*As a manuscript*

Romanova Anna Sergeevna

# DEVELOPMENT OF MANAGEMENT SYSTEMS USING ARTIFICIAL INTELLIGENCE SYSTEMS AND MACHINE LEARNING METHODS FOR BOARDS OF DIRECTORS

1.2.1 – Artificial Intelligence and Machine Learning

## DISSERTATION

for the purpose of obtaining academic degree of

Candidate of technical sciences

Academic advisor

Molchanov Evgeny Gennadievich,

candidate of physical and mathematical sciences,

associate professor

Moscow 2025

# Table of contents

## Introduction

**Relevance of the research topic. Due to the rapid development of artificial intelligence technologies, a paradigm shift** in corporate management is currently taking place: from the decision support mode, artificial intelligence systems (hereinafter AI) are moving to the decision management mode. The emergence of autonomous AI systems for corporate management creates a new working environment, as companies are forced to reconsider traditional management roles in order to incorporate autonomous AI systems into business processes [1].

The WEF 2015 report "Deep Shift: Technology Tipping Points and Societal Impact" predicted that by 2026, the first AI system would sit on a corporate board of directors [2]. The first official announcement of an AI system sitting on a board of directors was published in 2014, when Hong Kong-based venture capital firm Deep Knowledge Ventures announced the appointment of VITAL (Validating Investment Tool for Advancing Life Sciences) to its board of directors [3].

Since 2014, several artificial intelligence systems have been officially appointed to senior positions in international companies. For example, the humanoid robot Mika is the CEO of the Polish company Dictador [4], Alicia T is a top manager of the Swedish company Tieto [5], Spock is an artificial intelligence system of the Hong Kong company Deep Knowledge Ventures [6], Aiden Insight is a non-voting observer on the board of directors of the International Holding Company ( IHC) headquartered in Abu Dhabi [7], Tang Yu is the CEO of the Chinese IT company Fujian NetDragon Websoft Co., Ltd. [8]. Some companies are moving away from implementing individual personalised AI systems and instead are creating multifunctional digital command centres [9]. As Alex P. Miller, a professor at the University of Southern California, notes, **a quiet revolution** is currently underway, characterised by the steady increase in automation of traditionally human-made decisions in organizations [10].

In analyzing the history of the development of autonomous corporate management systems, it is necessary to take into account the concept of **technological singularity**, which is used in analyzing the development of machine artificial intelligence. Technological singularity is presented as an explosion of ever higher levels of intelligence, since each generation of machines, in turn, creates more intelligent machines [11]. Technological singularity populariser Vernor Vinge argues that if a technological singularity can happen, it will, because advances in automation are so compelling that passing laws or having customs prohibiting such things simply ensures that someone else will get them first [12]. Based on the theory of technological singularity, companies that introduce effective AI-based management will be more efficient and competitive, and accordingly, they will have the resources to implement an even more effective management system, which will again enhance their efficiency and competitiveness. For countries, regions, and companies with a shortage of skilled human capital autonomous AI systems can prove to be an effective tool, equalising or providing additional chances for such countries and companies in competition in the global market.

As many modern philosophers have noted, technological singularity contains enormous potential dangers: the end of humanity, an arms race between warring machines, the possibility of destroying the planet. And if there is even a small chance that a singularity will occur, we should think about what forms it might take and whether there is anything we can do to influence the outcome in a positive direction [11].

The main dilemma is that the development of information technology is much faster and more effective than the emergence of new approaches in the field of legislation and ethics. At present, there is no mandatory requirement that any scientific discovery, patent application, technical solution be accompanied by appropriate legislative and ethical justification. Such an infantile approach was still possible while humanity was playing with childish inventions, but will not work at the level of super technologies. As soon as the first stable technology for creating autonomous control systems appears, the process of technological singularity will begin. Some researchers consider cyborgization as a panacea or the only remedy for the singularity of machine intelligence [11-12]. But the singularity of cyborgization is subject to the same problems as the singularity of machine artificial intelligence.

A **"speed explosion"** [11] is also appropriate here to consider. The arguments for a "speed explosion" begin with the familiar observation that computer processing speed doubles at regular intervals. Faster processing will then lead to faster designers and even faster design cycles [11]. While modern society has had the opportunity to spend centuries creating effective corporate legislation, the rapid development of AI and big data technologies no longer provides that time. Successful implementation of autonomous AI systems for corporate governance requires a paradigm shift in which new legal and ethical technologies are developed simultaneously with new information technologies.

Another dilemma is the attempt to extrapolate legislation created for humans to autonomous AI systems. As Vernor Vinge rightly points out, most speculations about superintelligence seem to be based on the weak superintelligence model [12]. However, the best guesses about the post-singularity world can be obtained by thinking about the nature of **strong superintelligence** [12]. The creation of effective algorithmic law is a necessary condition for the development and implementation of autonomous AI systems that work for the benefit of humanity. The basis for the development of algorithmic law should be a full understanding of the fact that social **systems** (collectives of people) and **technical systems** (autonomous AI systems) are different in nature, and for the effective implementation of autonomous AI systems require the creation of a separate type of law — **algorithmic law**.

**Research area.** The following aspects are considered within the framework of the study:

- Methods and technologies for searching, acquiring and using knowledge and patterns, including empirical ones, in artificial intelligence systems. Methods and means of using expert knowledge.

- Formalisation and setting of control tasks and (support) decision-making based on artificial intelligence and machine learning systems. Development of control systems using artificial intelligence systems and machine learning methods, including control of robots, cars, etc.

- Development of specialized mathematical, algorithmic and software support for artificial intelligence and machine learning systems. Methods and means of interaction between artificial intelligence systems and other systems and a human operator.

- Ethical issues associated with the creation and implementation of AI systems, including modeling of expected social and economic consequences.

- Development of "strong AI", including the formation of a conceptual base and elements of mathematical formalism necessary for constructing an algorithmic apparatus.

- Development of "trusted" AI-class systems, including problems of forming test samples of precedents, reliability, stability, retraining, etc.

- Methods and means of forming arrays of conditionally real data and precedents necessary for solving problems of artificial intelligence and machine learning.

**The degree of development of the research topic.** The theoretical and methodological basis of the dissertation was the works of researchers in the fields of corporate law, artificial intelligence, machine learning, unmanned vehicles, robotics, game theory, ethics and philosophy of consciousness, as well as applied work on the application of artificial intelligence and machine learning technologies.

Significant contributions to the justification of the possibility of autonomous AI systems for corporate management were made by Meir Dan-Cohen, Shawn Bayern, Lynn LoPucki, John Armour, Horst Eidenmuller, Martin Petrin, Florian Moslein, Gian Mosco, Luca Enriques, Dirk Zetzsche, and Herbert Simon.

The theoretical aspects of the development of the foundations of algorithmic law were studied by Gottfried Wilhelm Leibniz, Pierre Simon Laplace, Stephen Wolfram, Paulus Meessen, Michael A. Livermore, Gemma Galdon-Clavell, Deborah Hellman, Wolfgang Hoffmann-Riem.

It should be noted that currently there is no legislation in any country in the world establishing the rules for the operation of autonomous AI systems in corporate governance. Existing autonomous AI systems operate practically in a legal vacuum. However, the basics of algorithmic law have already been fragmentarily formed, especially in those areas that are directly specified in the current legislation: non-discrimination, fairness, etc. They are mainly presented in the form of research articles and open-source libraries that provide various types of metrics, methods for detecting bias in data and algorithms for mitigating bias: AIF 360 [13], FA*IR [14], FairML [15], etc. Using these libraries, it is already possible to form a tiny model **of a digital director** for making management decisions in the field of fairness and non-discrimination. The development of effective legislation based on algorithmic law is one of the main prerequisites for the successful implementation of autonomous artificial intelligence systems for corporate management purposes.

The significant importance for the theory and practice of developing the foundations of ethics for artificial intelligence systems, unmanned vehicles, and anthropomorphic robots are the works of Patrick Lin, Edmond Awad, Azim Shariff, Jean - François Bonnefon, Iyad Rahwan, Jason Millar, Frederik J. Zuiderveen Borgesius and others.

The most striking example of the lack of a dedicated operational context for autonomous AI systems is the recognition of the robot Sophia as a citizen of the Kingdom of Saudi Arabia. Having the status of a citizen means that the robot operates in the same operational context as other citizens — people. Saudi Arabia is known for still holding strong religious and conservative values and still classifies Saudi women as second-class citizens, so it seems at best odd that the Kingdom would grant official citizenship status to a non-human creature that resembles on a woman [16]. Uncontrolled **legal capacity of robots** in the same operational context as ordinary citizens could lead to far-reaching and unpredictable social consequences.

However, even in the absence of a legislatively enshrined principle of a dedicated operational context for AI systems, developers are trying to form it themselves. These attempts are still fragmentary and manifest themselves in the most obvious areas (in particular, in the field of generative AI), and are formalised in the form of various tools and methods for developing responsible generative AI. Such tools include restrictions on the generation of malicious or toxic content, generation of watermarks, etc. [17].

Nowadays, synthetic data is already widely used in many various areas: from basic computer vision tasks to full-scale simulated environments for autonomous driving, drones, and robotics [18]. However, at the moment, synthetic data is not yet used with maximum efficiency in terms of the ethics and legitimacy of machine learning models. Nevertheless, initial research into creating tools for generating fair synthetic data is already underway. This includes, for example, FairGAN: Fairness-Oriented Generative Adversarial Networks [19], DECAF: Generating Fair Synthetic Data Using Causal - Aware Generative Networks [20] and others .

The main approach to planning system actions in the development of autonomous AI systems is currently to focus on algorithm enumeration [21]. However, a simple enumeration of algorithms does not guarantee that the AI system will achieve its goal. Autonomous AI systems should be able to calculate an effective strategy for achieving the set goals. John von Neumann, Oskar Morgenstern, John Forbes Nash, Abraham Wald, Leonid Gurvits, and others made a significant contribution to the creation and development of mathematical modeling of decision-making in conflict situations. Open source libraries already exist for strategy modeling and simulation: Nashpy [22], Gambit [23], etc. However, they require adaptation for use by autonomous AI systems.

The active use of artificial intelligence algorithms has led to extensive research in the field of AI explainability. Numerous open-source algorithms and libraries for studying AI explainability are already available to interested experts (for example, AIE 360 [24] and others).

Currently, the practical implementation of autonomous AI systems for corporate management is being carried out by companies such as ADNOC, Dictador, IHC, Deep Knowledge Ventures, Tieto, Fujian NetDragon Websoft Co., Ltd. and others significantly outstrips the scientific substantiation of methods for developing and implementing such systems. Poor-quality development and implementation of AI systems for corporate management carry not only economic risks for interested companies and their shareholders, but also existential risks for all of humanity.

**Objectives and tasks of the study.** The aim of the study is to theoretically and methodologically substantiate methods for developing and implementing autonomous AI systems for corporate management in order to facilitate the implementation of autonomous AI systems for use in industrial settings and to provide researchers with a general framework for the development and implementation of such systems for evaluation.

The set goal determined the need to set and solve the following tasks:

1) To identify established patterns in the development of autonomous AI systems for corporate management.

2) To develop and justify a taxonomy of autonomous AI systems for corporate management.

3) To propose, justify and test a reference model for the development and implementation of autonomous AI systems for corporate management.

4) To substantiate and test a methodology for developing algorithmic law for autonomous AI systems for corporate management.

5) To justify and test a methodology for creating a dedicated operational context for autonomous AI systems for corporate management.

6) To justify and test a methodology for training autonomous AI systems for corporate management based on synthetic data.

7) To substantiate and test a methodology for calculating the strategy of autonomous AI systems for managing corporations based on game theory.

8) To justify and test a methodology for developing an interface for autonomous AI systems for corporate management.

9) Develop and test a continuous process for making legitimate and ethical management decisions by autonomous AI systems.

**The object of the study** is the corporate governance system of commercial companies.

**The subject of the research** is the socio-economic and legal relations arising between technical (autonomous AI systems) and social systems (groups of people) in the process of making management decisions by boards of directors using AI technologies.

**The scientific novelty of the study** lies in solving a promising scientific problem: creating theoretical and methodological foundations for the development and implementation of autonomous AI systems for corporate management.

**The theoretical significance** of the study lies in the development of the theory of artificial intelligence in terms of proposing a reference model and developing and implementing autonomous AI systems based on the synthesis of computational law, dedicated operational context, controlled generation of synthetic data, game theory (used to calculate the strategy for achieving goals by an AI system), explainable AI technologies and machine learning algorithms.

**The practical significance** of the study is due to the fact that the proposed reference model for the development and implementation of autonomous AI systems for corporate management can serve as a methodological basis for the development and industrial implementation of autonomous AI systems for commercial and civil purposes.

**Methodology and research methods.** In accordance with the stated goal and objectives, general scientific methods of corporate law theory, artificial intelligence, machine learning, unmanned vehicles, robotics, game theory, ethics and philosophy of consciousness were used. Tabular and graphical methods of data visualization were applied.

**The theoretical and practical basis of the study** was the analytical data of the World Economic Forum, Deloitte, KPMG, Accenture and other companies, as well as materials on the development and implementation of existing autonomous AI systems introduced by ADNOC, Dictador, IHC, Deep Knowledge Ventures, Tieto, Fujian NetDragon Websoft Co., Ltd. and others.

**Provisions submitted for defence:**

1) Patterns in the development of autonomous AI systems for corporate management have been identified.

2) The taxonomy of autonomous AI systems for corporate management is proposed and substantiated. The emergence of hybrid systems as a result of the combination of multifunctional digital factories and personalised virtual systems or humanoid robots is substantiated.

3) The reference model for the development and implementation of autonomous AI systems is proposed and substantiated based on the synthesis of computational law, a dedicated operational context, controllable generation of synthetic data, game theory (used to calculate the strategy for achieving goals by an AI system), explainable AI technologies and machine learning algorithms.

4) The necessity is substantiated, a methodology for developing algorithmic law for autonomous AI systems for managing corporations is proposed and tested.

5) The necessity is substantiated , a methodology for creating a dedicated operational context for autonomous AI systems for corporate management is proposed and tested.

6) The methodology for training autonomous AI systems for corporate management based on synthetic data has been proposed and tested.

7) The methodology for calculating the strategy of autonomous AI systems for managing corporations based on game theory is proposed and tested.

8) The methodology for developing an interface for autonomous AI systems for corporate management has been proposed and tested.

9) The continuous process of making legitimate and ethical management decisions by autonomous AI systems that combines computational law, dedicated operational context, controlled generation of synthetic data, game theory, explainable AI technologies, and machine learning algorithms is presented.

**Degree of reliability, testing and implementation of research results.** The main results of the work were reported at the following scientific conferences and seminars:

1) on OpenTalks.AI 2023 (RA, Yerevan, March 6-7, 2023);

2) at the 65th All - Russian Scientific Conference of MIPT (RF, Dolgoprudny, April 3–7, 2023);

3) at the All-Russian interdisciplinary seminar "Artificial Personality Project" of the Scientific Council under the Presidium of the Russian Academy of Sciences on the Methodology of Artificial Intelligence and Cognitive Research (NSMI RAS) (RF, Moscow, April 26, 2023);

4) at the First Russian Conference "Consciousness, Body, Intelligence and Language in the Era of Cognitive Technologies" (RF, Pyatigorsk, September 28–30, 2023);

5) at the round table "Robophilosophy: an interdisciplinary approach" (Congress "Russian Engineer", conference "Bionics - 2023") (RF, Moscow, November 2, 2023);

6) at the conference "HES. Digital. Law" (RF, Moscow, February 15, 2024);

7) at the 66th All - Russian Scientific Conference of MIPT (RF, Dolgoprudny, April 1–6 , 2024);

8) at the 12th international conference "Artificial Intelligence and Natural Language" (RK, Alma - Ata, April 24-25, 2024);

9) at the CGI Russia round table "Digitalisation of corporate governance: will enter AI board of directors ?" (RF, Moscow, June 18, 2024);

10) at the III International Scientific and Practical Conference "Digital Technologies and Law" (RF, Kazan, September 20, 2024);

11) at the Ninth All-Russian scientific and practical conference with international participation "Artificial intelligence in solving urgent social and economic problems of the 21st century" (RF, Perm, October 17-18, 2024);

12) at the round table "AI in business, industry and legal issues" (RF, Moscow, October 23, 2024);

13) at the international interdisciplinary conference "Philosophy of Artificial Intelligence" (NSMI RAS) (RF, Moscow, October 23–24, 2024);

14) at the XXVI International Scientific and Technical Conference "Neuroinformatics - 2024" (RF, Moscow, October 21–25, 2024);

15) at the II All-Russian Congress with international participation "Russian Engineer" (RF, Moscow, October 30 - November 1, 2024);

16) at the First International Conference Artificial Intelligence Research (AIR) (UAE, Dubai, December 10–12, 2024);

17) at the 13th international conference "Artificial Intelligence and Natural Language" (RF, Novosibirsk, April 18-19, 2025).

The materials of the dissertation are used in the development of the project "iBoard - Autonomous Board of Directors". The iBoard project was recognised as the best project of the PHYSTECH.AERO (RF, Moscow, MIPT, April-June 2023). The iBoard project was also presented at International competition of the best technological practices BRICS Solutions Awards 2024 (RF, 2024) and at the National Network Accelerator "Archipelago 2024") (RF, Yuzhno-Sakhalinsk, July 10-26 , 2024). Testing The results of the study are confirmed by relevant documents.

**Publications.** The dissertation materials have been published by the applicant in sufficient detail in the following works:

the Problems of Developing Autonomous Artificial Intelligence Systems for Corporate Management) // Consciousness, Body, Intelligence, and Language in the Era of Cognitive Technologies: Abstracts of the First All-Russian Conference "Consciousness, Body, Intelligence, and Language in the Era of Cognitive Technologies (MBIL-2023)", September 28–30, 2023, Pyatigorsk State University. — 2023. — P. 177-178.

12) Romanova A. S. Dedicated operational context as a basis for the implementation of autonomous artificial intelligence systems // Proceedings of the 66th All - Russian Scientific Conference of MIPT , April 1-6, 2024. Applied Mathematics and Computer Science . — 2024. — P. 268–270.

13) Romanova A. S. Modeling of autonomous artificial intelligence systems for corporate management // Digital technologies and law: collection of scientific papers of the III International scientific and practical conference (Kazan, September 20, 2024) . - 2024. - V. 3. - P. 281-288.

14) Romanova A. S. Algorithmic law as a basis for the implementation of autonomous systems for corporate management // Artificial intelligence in solving urgent social and economic problems of the 21st century: a collection of articles based on the materials of the Ninth All-Russian scientific and practical conference with international participation (Perm, October 17-18, 2024). - 2024. - P. 47-51.

15) Romanova A. S. Main types of autonomous systems for corporate management // Collection of abstracts of the II All - Russian Congress with international participation "Russian Engineer". - 2024. - P. 170-171.

**Structure and volume of the dissertation** are determined by the purpose, objectives and logic of the research. The dissertation consists of an introduction, three chapters, a conclusion, a glossary of terms, a bibliography consisting of 162 titles, and two appendices. The text of the dissertation is presented in 140 pages, contains 18 tables and 13 figures. The research code and the artificial data set are available in the repository: https://github.com/iboard-project.

# Chapter 1

## Theoretical and methodological foundations of the development of autonomous artificial intelligence systems for corporate management

The materials of this chapter are based on works [1, 4, 6] from the list of publications.

### 1.1 Modern approaches to the use of autonomous artificial intelligence systems for corporate management

The main prerequisites for the development of systems for the complete automation of management decisions made at the level of the board of directors (hereinafter BoD) are formed in the field of corporate law, machine learning, compliance with non-discrimination rules, transparency and accountability of decisions made and algorithms applied.

**Development of legal framework for the use of artificial intelligence systems for corporate management.** Methodological approaches in the field of corporate law for the creation of an "autonomous director" began to form at the end of the 20th century. University of California professor Meir Dan-Cohen proposed the concept of a fully **automated corporation** in 1970 — **"personless corporation"** [25]. According to Dan-Cohen, replacing corporate management with computers will have a minimal effect on the operating activities and legal status of the corporation. Dan-Cohen, however, points out that a necessary condition is precisely the successful transition to automated decision-making. If successful, the legal status of the corporation will not change. According to Dan-Cohen, a fully automated corporation will easily pass a kind of Turing test for corporations: profits will be reinvested, friendly politicians and cultural events will be sponsored in a timely manner [25]. Dan-Cohen points out that an autonomous corporation will have to have rights. Thus, a solution to the problem is necessary: can a fully autonomous corporation have rights? Dan-Cohen predicts that the answer will be positive [25].

Dan-Cohen's theoretical ideas have been further developed in the 21st century in connection with the successful development of information technology. A practical approach was demonstrated by Florida State University professor Shawn Bayern in a series of articles devoted to **autonomous companies** [26]. Bayern showed that anyone could give legal personality to an autonomous computer algorithm by placing it under the control of a limited liability company. Professor Lynn LoPucki of the University of California, Los Angeles , in continuation of the work of Shawn Bayern, introduced the concept of **an algorithmic company** (an **"algoritmic entity"**), where an algorithm will fulfill all the rights and obligations of a legal entity [27]. LoPucki also showed that a company created under the laws of the state of Delaware does not necessarily have to be managed by an ordinary manager, but can be managed by an artificial person [27]. LoPucki analyzed the current legal readiness for the emergence of algorithmic companies in several jurisdictions. The analysis revealed that even without modifying current corporate legislation, legal structures for the operation of algorithmic companies are already possible [27].

Along with the analysis of practical possibilities, the applied terminology and tools for creating an autonomous BoD have also evolved. If Dan-Cohen used the term **"personless corporation"**, then in 2019, John Armour, Professor of Law and Finance at the University of Oxford, and Horst Eidenmuller, Professor of Commercial Law at the University of Oxford, introduced the term **"autonomous / unmanned (self-driving) company"** – by analogy with a self-driving car [28]. Also in 2019, University of California, Los Angeles, professor Martin Petrin introduced the concept **of a company without leaders** ("**leaderless entity"**) and developed the concept **of the algorithmic company** [29]. There are already algorithms that can exist autonomously. The simplest example is computer viruses. Petrin suggests that advanced forms of such algorithms can run a business on their own. Petrin considers the question of whether there would be a need for corporate management in this case. Petrin proposes the concept of new business structures that would function without leadership in the traditional sense. Petrin gives the example of ConsenSys, a software development company specialising in applications for the Ethereum blockchain platform [29].

Professor Petrin also introduces the concept of **a fused board**: where the various roles and contributions previously provided by a team of human directors are incorporated into a single software or algorithm whose performance will outperform today's human-provided management [29]. Petrin believes that in the future, AI systems will also replace managers and sub-managers, and these developments will ultimately make separation of board and management is obsolete and will lead to **fused management** of corporations, with companies being completely managed by one division of AI [ 29 ].

Florian Moslein, a professor at the University of Marburg, associates the transition to autonomous AI management primarily with the emergence of **decentralized autonomous organizations (**DAOs), which are governed according to rules encoded in the form of computer programs (so-called smart contracts); these rules, as well as records of their transactions stored in the blockchain, mean that they can operate completely without human intervention [30]. Moslein believes that for a mixed BoD or mixed management to work effectively, it will be necessary to introduce rules for distributing decision-making rights between humans and machines [30]. Professor Moslein uses the concept of a **robo-director** [30]. Gian Mosco, a professor at the Guido Carli University, by analogy with Elon Musk's term "robotaxi" proposes the term **"roboboard"** which refers to a board of directors consisting of artificial directors [31].

In 2019, Oxford University professor Luca Enriques and University of Luxembourg professor Dirk Zetzsche introduced the general term **CorpTech** [32]. In Enriques and Zetzsche's formulation, CorpTech includes all solutions related to corporate governance in a broad sense, including tools for setting compensation, identifying candidates for senior positions in the organization, maintaining relations with investors, procedures for corporate voting and internal work of the board of directors, risk management and improving the effectiveness of the board of directors function [32].

One of the generally recognised global standards of corporate governance is the G20/OECD Principles of Corporate Governance [33]. The G20/ OECD Principles of Corporate Governance define the key elements of the corporate governance structure and offer practical guidance for their application at the national level [33]. The G20/ OECD Principles do not establish a requirement for the duties of a director to be performed by an individual. For example, in modern corporate practice, there is a concept of a corporate director, when legal entities act as directors of a company [34]. The G20/ OECD Principles consider the definition of a person who can perform the duties of a director to be the prerogative of the local legislation of each country. Thus, if the local legislation allows the use of an AI system as a director, and the available technologies support the actual performance of the necessary functions of a director, then the corporation can legitimately appoint an autonomous director to the Board of Directors.

**Using artificial intelligence technologies to manage corporations**. Back in 1965, one of the pioneers of artificial intelligence, Professor Herbert Simon of Carnegie Mellon University in Pittsburgh, formulated the concept of **an autonomous factory** and the concept of **an autonomous** **corporation** [35]. Simon predicted the automation of almost all clerical work, which actually happened, and also predicted the emergence of **automated executives**. In his view, "complex information processing systems" would be able to replace people in tasks where it was necessary to "think and make decisions". Top managers, in Simon's understanding, would be engaged only in those tasks where emotional interaction with workers was necessary. Simon expected that all the major changes would have already occurred by 1985 [35].

According to an international consulting company Accenture sees the transition of AI systems from a passive to an active role in company management occurring in three main stages:

1)  AI as an assistant - creates scoring cards, maintains reports, takes notes, manages the schedule, monitors the environment, tracks the implementation of decisions made;

2)  AI as an advisor - answers questions, develops scenarios, generates options, leads meetings, analyses team behaviour, recommends roles in the team;

3)  AI as an actor - evaluates options, makes decisions, conducts budgeting and planning, evaluates team dynamics, changes the status quo [36].

In accordance with the G20/ OECD Principles, the main functions that the board of directors should perform are: determining corporate strategy, monitoring the effectiveness of company management, selecting key management officials, determining risk management policies and procedures, etc. [33]. At the current level of technological development, most corporate management functions are either in the process of automation or have already been automated. Commercial companies and researchers are actively developing in the field of further automation of management decisions.

An AI system that acts as an assistant in executive meetings was unveiled back in 2018 at the WEF meeting in Davos [29]. It's called Einstein, which SalesForce CEO Marc Benioff uses in his weekly staff meetings. Benioff gave an example of how Einstein was able to assess the situation more objectively than a highly trained manager when discussing plans: "I have 30 or 40 senior managers around the table," he ( Benioff ) said. "And we figure out how we're doing when we look at all this analysis". Benioff said that after he's worked in his office with the executives, he then quizzes his artificial assistant. "I ask Einstein, 'I've heard what everyone's saying, but what do you really think?'" A Salesforce executive said he's been using the technology for a year, and Einstein recently questioned a European employee's reports. "And Einstein said, 'Well, I don't think that executive is going to be successful, I 'm sorry.' " [37].

Suppliers of existing ERP systems are also developing special modules for automation of the work of the boardroom, in particular, SAP Digital Boardroom. The current version already supports the following functions: intelligent meetings (supports real-time analysis, provides interactive data visualizations and adaptive layouts), instant analytics based on data, planning and modeling, etc. [38].

Machine learning researchers have convincingly demonstrated that AI algorithms successfully cope with many board tasks: determining the optimal dividend policy [39], electing board members [40], supporting voting decisions at the annual general meeting of shareholders [41], etc. To solve management problems, AI algorithms are used separately (to perform a separate management function) or are combined into multifunctional command centres.

**Transition from automation of corporate management to autonomy.** The main question at the moment is when and how the qualitative step from automation of corporate governance to **autonomy** will occur. A necessary condition for the autonomy of AI systems is compliance with the same requirements of non-discrimination, transparency, and accountability as for ordinary ("natural") directors.

Modern researchers propose three main approaches to managing the risks that arise when using artificial intelligence models: transparency, explainability, and accountability [42]. Researchers identify several stages of AI system development where the main bias errors may occur: input data, training, and programming [42]. Several approaches are currently being developed that aim to develop methods for eliminating errors in AI systems. These approaches are mainly reflected either in government regulation of the process of developing and using AI systems, or in the use of certain technologies.

In particular, the European Union is attempting to address the black box problem and the resulting transparency problem by providing EU citizens with the so-called "**right to explanation**" [43]. It is assumed that the introduction of the "right to explanation" will force developers to structure algorithms and systems differently to provide a higher degree of transparency [43]. Regulations that require certain AI systems to explain themselves already exist and have been evolving for several decades. In particular, these are credit rating systems and GDPR personal data systems [44].

In 1970, the Fair Credit Reporting Act was passed in the United States, followed by the Equal Credit Opportunity Act in 1974 [44]. This law prohibited discrimination in credit decisions based on race, color, religion, national origin, sex, marital status, age (for adults), and receipt of income from public assistance [44]. However, in the process of providing explanations for credit decisions, it turned out that access to a model is not the same as understanding the model. It is believed that modern credit systems are not based on complex algorithms precisely because of the initial existence of a legislative requirement for an understandable explanation [44]. There is a clear trade-off between the performance of a machine learning model and its ability to make explainable and interpretable predictions [45].

Technologically, the "right to explanation" for AI systems is realised through the area of **Explainable AI (XAI)** – explainable or transparent AI that people can easily understand. Explainable artificial intelligence (XAI) is a field that develops new methods that explain and interpret machine learning models and has gained enormous popularity in recent years [45].

Another widely used method is independent qualified testing, such as the Facial Recognition Vendor Test (FRVT) conducted by the US National Institute of Standards and Technology. Algorithms are submitted for testing in compiled form and tested as a "black box" with a C++ testing interface. The test developers believe that the main result of the assessment is that significant gains in accuracy have been achieved over the past five years (2013–2018) [46]. Since not all companies are willing or actually able to disclose and explain how the algorithms of the AI system they have created work, the option of independent testing, auditing and certification is a mechanism to ensure that the principles of non-discrimination, transparency and accountability of the AI system are met.

**Responsibility for decisions made by AI systems.** With such a significant transfer of powers to autonomous systems, the question of responsibility for the decisions made by such systems inevitably arises. Professor Petrin suggests that the simplest option for distributing rights and responsibilities would be the already applied system of "**manufacturer responsibility**" [29]. This approach is already being implemented in practice. For example, when discussing the Artificial Intelligence Act, the European Commission declares that responsibility for the operation of AI systems may be assigned to the supplier, importer, distributor, operator or user [47]. At the moment, the European Commission does not consider AI as a legally competent person. In the terminology of the Act, AI systems are considered as a technical tool, a product entering the market, and responsibility for its use is assigned to all participants in the value chain of such a product and its users.

The European Commission acknowledges that autonomous behaviour may negatively impact a number of fundamental rights enshrined in the EU Charter of Fundamental Rights [47]. However, autonomous systems for managing companies are not even described in the Artificial Intelligence Act. Systems that have some initial functions for managing companies are classified as high-risk AI systems: if the AI system is designed to make decisions on the promotion and termination of employment contracts, to distribute tasks, as well as to monitor and evaluate the performance and behaviour of individuals in such relationships. Providers of such systems are subject to additional requirements for risk monitoring and registration of the systems they develop [47].

Despite the variety of attempts to guarantee the ethics and legitimacy of decisions made by autonomous AI systems, at the current level of development of human society and technology, this problem remains unsolved. This study proposes approaches that will allow the development of **ethical** and **law-respecting** autonomous AI systems for corporate governance.

### 1.2 Main types of autonomous artificial intelligence systems for corporate management

The main types of autonomous AI systems currently being developed to manage corporations are multifunctional digital command centres (hereinafter referred to as DCC), personalised virtual systems, and anthropomorphic robots (figure 1).

Digital command centres are currently being implemented by many commercial and non-commercial organizations. They focus primarily on **big data-based** decision making and do not have social interaction functions with people. Companies that value **social interaction** implement personalised systems in the form of virtual agents or humanoid robots. **Hybrid systems** are also expected to emerge by combining multifunctional digital command centres (for big data processing) and personalised systems (as an interface with social interaction functions - in the form of a virtual agent or humanoid robot).

**Digital command centres.** Many companies are moving toward automating decision making at the top management level not by formalising a structure such as a key decision center for the company, but by creating digital command centres in practice . Command centres have been around for a very long time; large organizations and governments have used them since the beginning of civilization . However, the technologies that enable command centres to perform their tasks have changed dramatically , and new skills and approaches are required to ensure success [48].

One modern example is the digital command center of the oil company ADNOC, which was commissioned several years ago. The Panorama digital command center integrates real-time information from more than a dozen subsidiaries and joint ventures (figure 2). The system also uses a smart analytics model and artificial intelligence to generate operational conclusions and recommendations [49].

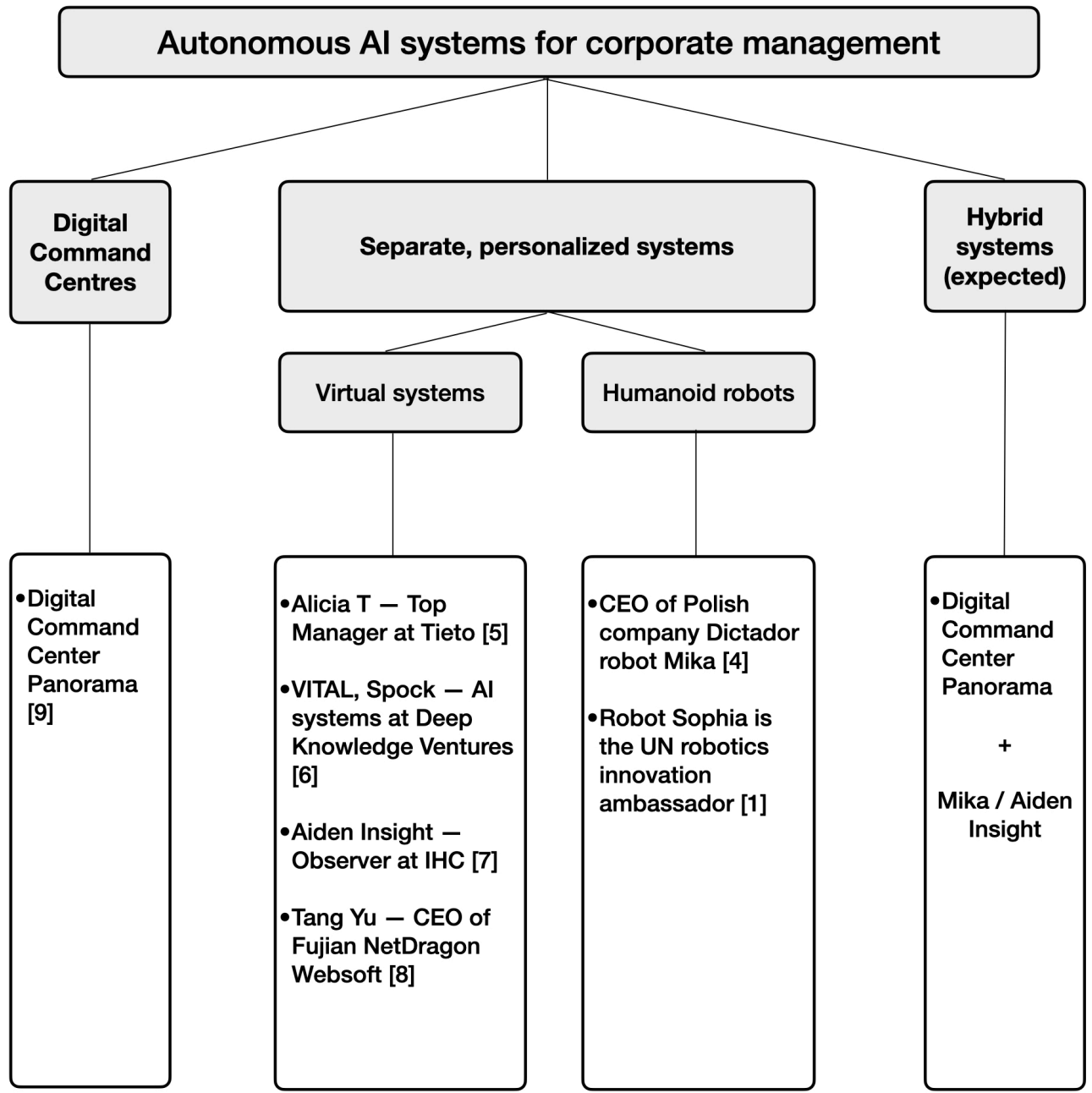

Figure 1 - The main types of autonomous AI systems for corporate management. Source: compiled by the author.

Panorama DCC demonstrates in practice how theoretical discussions about the change in the top management paradigm in the AI era are implemented: reduction of agency costs, transparency and accountability of information, real-time data tracking, etc. Panorama DCC uses digital twins technology for monitoring, and blockchain technology is used to record and store information. In essence, Panorama CCC is a digital factory that processes information and makes recommendations within a company with a turnover of more than US $60 billion [49]. Official strategy of the oil

company ADNOC (UAE) until 2030 envisages that Panorama's digital command center will eventually be connected to customers and investors, providing continuous integration between stakeholders [50]. In 2021, ADNOC won an industry technology award for its digital command center [9], marking recognition and further dissemination of similar technologies among multinational companies.

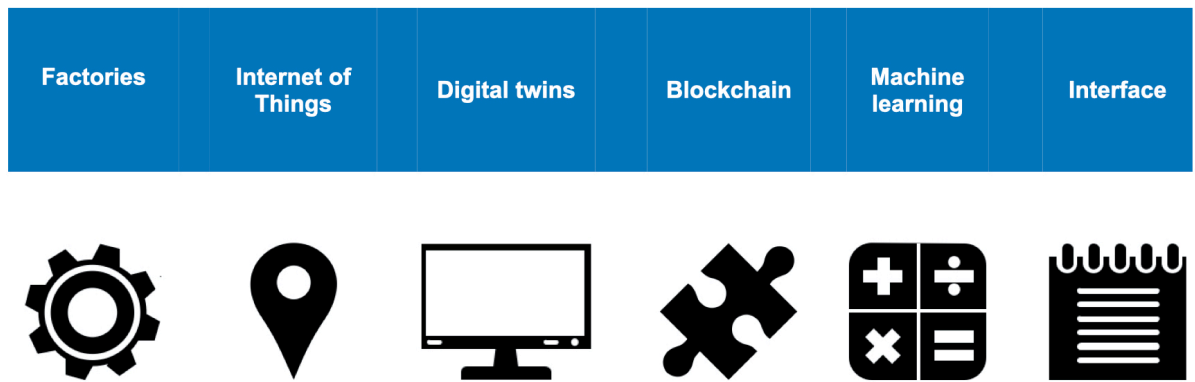

Figure 2 — Structure of digital command centres using the example of the Panorama Digital Command Center. Source: compiled by the author.

One of the underlying technologies that enables real-time insights across the entire company is digital twins. The technology emerged at the beginning of this century and aimed to create a digital model of a physical system before it was built. The technology then combined with IoT sensors and as a result it became possible to add more data to enrich the digital twin, collecting environmental data such as location and configuration, as well as more general information such as service records, financial models, etc. [51]. The result is that organizations have a digital construct that knows everything about the design, construction or production, past and current operations and maintenance of the physical system. By adding intelligence in the form of analytics, models and other algorithmic methods such as machine learning, organizations can begin to receive predictions and early warnings faster than ever before [51].

Another technology that has improved the transparency and credibility of management decisions is **blockchain technology**. Modern blockchain technology became famous in 2008 when it was used to create the cryptocurrency Bitcoin [52]. Blockchain serves as an immutable ledger that allows transactions to be conducted in a decentralized manner. Thus, when making a management decision that is based on information recorded in the blockchain, one can expect that the information is credible and transparent.

The third, important component that has enabled the creation of **an "operational digital brain"** on an industrial scale is **machine learning** technology. The simplest formulas used in machine learning emerged many years ago, but machine learning technology has advanced most significantly in the last two decades [53]. Recent progress in machine learning has been driven both by the development of new learning algorithms and theory, as well as the ongoing explosion in the availability of online data and low-cost computing [53].

**Anthropomorphic robots.** Currently, the most well-known achievements of two humanoid robots are made by CEO Mika and the robot ambassador Sophia [54]. Mika is a member of the board of directors, oversees the Arthouse Spirits DAO project and facilitates communication with the DAO community on behalf of the Polish company Dictador [55]. Mika realises the benefits of humanoid robots by providing an exclusive opportunity for members of the DAO community to meet and spend time with her [1]. There is a point of view that CEOs — humanoid robots, free from human biases and emotions, have the ability to consistently uphold ethical standards, encouraging ethical and social responsible behaviour in the workplace [1].

The robot Sophia converts written text into speech, while a more advanced speech generation system, the intelligent chatbot, is used for interactive conversations, using Google algorithms to understand human queries, search for answers in a database, and generate short responses [1]. CEO Mika, an advanced artificial intelligence robot developed by Hanson Robotics, is an improved version of its prototype sister, Sophia [ 1].

**Virtual agents.** Virtual agents share many of the same anthropomorphic benefits as humanoid robots, but do not have a physical body.

The most prominent modern implementation of a virtual agent is the CEO of a Chinese gaming company, Tang Yu. NetDragon Websoft, a gaming company headquartered in Fuzhou, has appointed a "virtual humanoid robot with artificial intelligence" named Tang Yu as CEO of its subsidiary Fujian NetDragon Websoft. Following the announcement of the appointment, the company's shares outperformed the Hang Seng Index, which tracks the largest companies listed in Hong Kong [56]. It is expected that Tang Yu will streamline processes, improve work quality, improve execution speed, and serve as a real-time data hub for analytical decision making and risk management. Also Tang Yu will realise the potential of virtual agents to work in the Metaverse [1].

**The emergence of hybrid systems.** A predictable stage in the development of autonomous AI systems for corporate management is the emergence of hybrid systems by combining multifunctional digital factories and an interface in the form of personalised virtual systems and / or humanoid robots (figure 3). Hybrid systems will be able to combine the capabilities of digital command centres to work with big data and the ability of anthropomorphic robots and virtual agents to communicate with people and for corporate leadership .

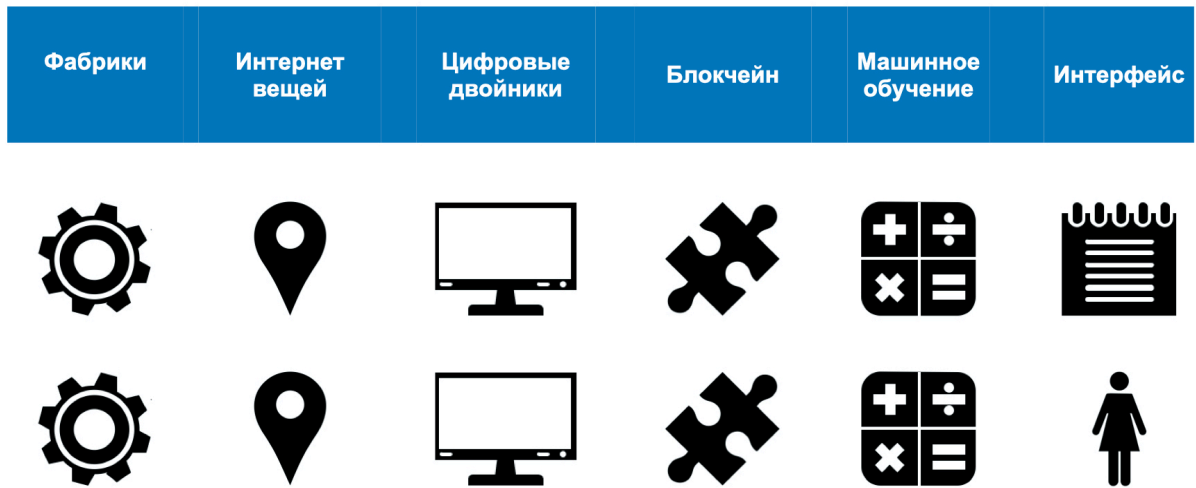

Figure 3 – The emergence of hybrid systems. Source: compiled by the author.

Multifunctional digital command centres seem to be a more promising direction of research than individual, personalised systems. Given the general trend towards digitalisation of production, it can be expected that multifunctional digital factories will emerge naturally in the process of digitalisation.

The competitiveness of personalised systems will largely depend on the possibilities of social interaction: either they will become interfaces of digital factories, or they will be replaced by systems with a more informative presentation of data and the process of their analysis.

The Internet of Things, digital twins and blockchain allow machine learning models to obtain information from the real world and avoid the risk of hallucinations that, for example, large language models are subject to. In addition, this model makes it possible to consider the hypothesis of **continuity of "machine consciousness"**: the blockchain can store all accumulated experience, and digital twins can store internal representations of the outside world.

## 1.3 Model for the development and implementation of autonomous artificial intelligence systems for corporate management

Despite the fact that the practical implementation of autonomous AI systems for corporate management is an active and promising area, the methodological, regulatory and legislative foundations for the creation of such systems are almost completely absent. To address these issues, a reference model for the development and implementation of autonomous AI systems for corporate management is proposed (figure 4). The proposed model for the development and implementation of autonomous AI systems is based on the synthesis of computational law, a dedicated operational context, controlled generation of synthetic data, game theory (used to calculate the strategy for achieving the goals of the AI system), explainable AI technologies and machine learning algorithms. The goal of the proposed model is to promote a deeper understanding of the necessary aspects of the development and implementation of autonomous AI systems, provide an open model for researchers and industry practitioners to share and compare their approaches and facilitate the transition to use in industrial settings .

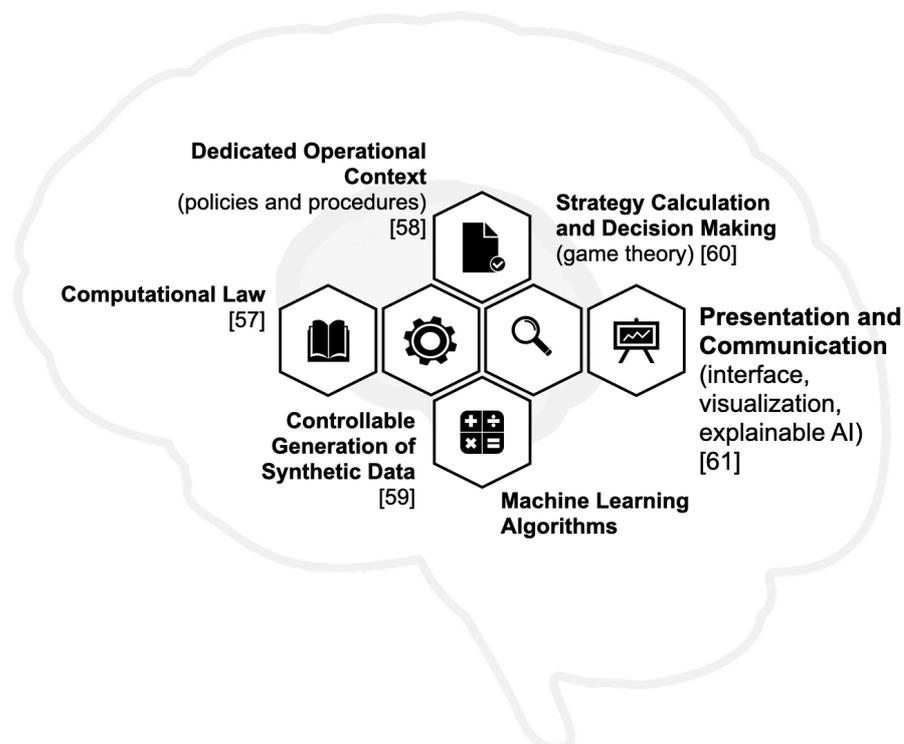

Figure 4 - The reference model for development of autonomous AI system for corporate management. Source: compiled by the author.

The overall structure of the model provides a link between the conscious choice and informed consent of all parties interested in the success of the company and the practical implementation of machine learning algorithms, computational law, synthetic data generation, game theory, and explainable AI technologies.

Computational (algorithmic) law is a necessary condition for the creation legitimate and ethical autonomous AI systems for corporate governance. The concept of law as a computation aims to reduce the law to a set of algorithms that can be automatically executed on a computer, smoothly transforming raw input data into legal conclusions. The need for algorithmic legislation was considered by famous mathematicians, in particular, Gottfried Leibniz [62] and Pierre Laplace [63], but at that time they did not have the necessary information and social technologies to implement their ideas. In the proposed model, computational law means an approach when legal aspects are reflected in local policies at the company level or at the federal legislation level using mathematical apparatus for unambiguous interpretation by autonomous AI systems.

A significant part of the operational context for autonomous corporate governance systems is the regulatory and legal environment within which corporations operate. In order to create a dedicated operational context for autonomous AI systems, the wording of local regulatory documents can be simultaneously presented in two versions: for use by people and for use by autonomous systems. In this case, the AI system receives a clearly defined operational context, which allows such a system to perform functions within the framework of the required operational qualities. The dedicated operational context makes it possible to create conditions for the joint work of people and autonomous systems: individuals act within the framework of codes, policies, and procedures intended for people. Autonomous AI systems act within the framework of codes , policies , and procedures intended for AI systems. The method of creating a dedicated operational context is widely used in legal practice, for example, when concluding bilingual contracts and other bilingual and multilingual documents. As a result, each party receives not a general operational context, but a dedicated one, the terminology of which it understands and within which it can effectively fulfil its obligations.

Controllable generation of synthetic data allows us to correct the mistakes and shortcomings of ordinary top managers and train autonomous AI systems to act more effectively, legitimately and ethically in the context of continuous development of human society, legal requirements, business technologies and business ethics. The proposed technique allows using synthetic data to train machine learning classifiers on almost any issues that may interest the company's board of directors. While actual data for training purposes may be unavailable, supervised generation of synthetic data allows us to reflect any nuances that the company needs.

In order for an autonomous AI system to achieve its goals, the autonomous system needs a strategy calculation methodology. Game theory is widely used both to analyze the effectiveness of interactions between directors and shareholders [64] and in making management decisions. A company can determine priority strategies for an AI system and monitor their implementation. A dedicated operational context for AI systems will allow for the effective formulation of game conditions for the AI system, as well as the discovery and optimization of possible game strategies.

To effectively interact with shareholders, investors, employees and colleagues, an autonomous AI system must have a user-friendly interface. Autonomous AI systems must be able to demonstrate that the decisions they make are logical, legitimate and ethical when making important management decisions. The explainable AI techniques discussed in modern literature (mainly text explanations and visualizations [65]) can be used by both digital command centres and personalised systems to present and communicate their decisions.

Using the proposed model to process data obtained from the Internet of Things, digital twins, and stored using blockchain technology will allow the creation of an ethical and law-abiding **operational digital brain** on an industrial scale (figure 5).

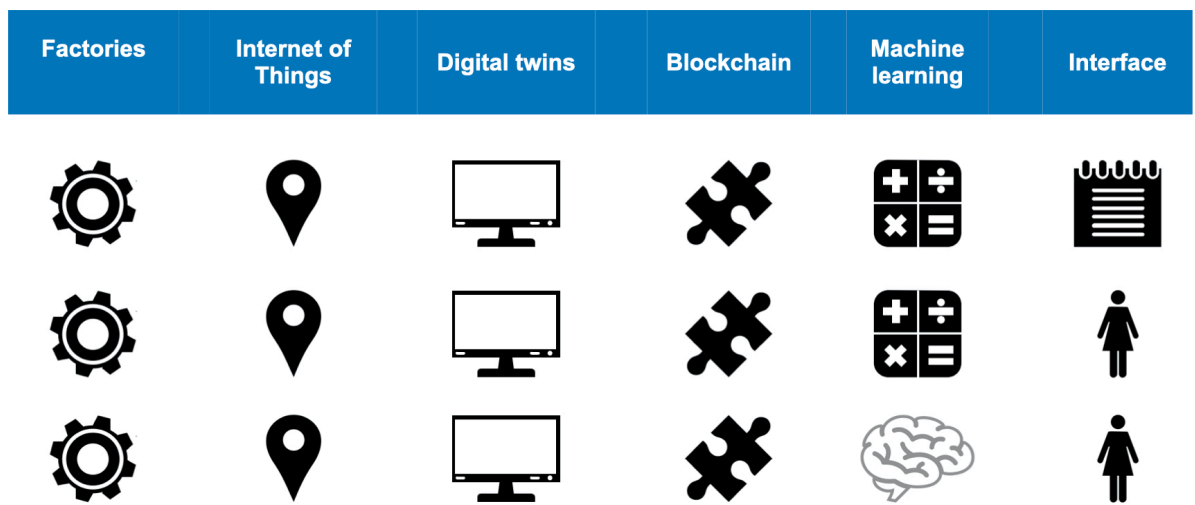

Figure 5 - The operational digital brain for corporate management. Source: compiled by the author.

**Chapter 2**

**Main components of a model for the development and implementation of autonomous AI systems for corporate management**

The materials of this chapter are based on works [ 2, 3, 4, 7, 9 ] from the list of publications.

**2.1 Computational law as a basis for the creation and use of autonomous artificial intelligence systems for corporate management**

One high-level question is whether it is worthwhile to engage **in law as computation** at all [62]. Directors who are natural persons and directors who are autonomous AI systems must comply with corporate governance requirements and standards established by law, authorized regulators and companies. The basic rules of corporate governance of a company are usually set out in internal policies and procedures. The G20/ OECD Principles of Corporate Governance indicate that well-formulated corporate governance policies play an important role in achieving the broad economic objectives of public policy [33]. However, codes, policies and procedures developed for individuals are not always fully applicable or in principle inapplicable to AI systems. As numerous legal researchers have noted, automation of legal norms can currently only exist thanks to **compromises** reached in translating laws in natural languages into code (executable files) in formal languages [66]. The need for a compromise is caused primarily by the fact that human decisions are **social constructs**, while decisions made by AI systems are **technical constructs** [67]. To resolve this contradiction, it is proposed to formulate the principles of making management decisions **in two versions**: in the usual version - for human decision-making, and in the algorithmic (mathematical) version - for decision-making by an AI system. In this way, mixed boards of directors (consisting of individuals and AI systems) will receive a high-quality environment for effective collaboration.

**A reference book of algorithmic terms** can serve as a minimum basis for the creation and development of computational law. Below, a methodology for creating a reference book (dictionary) for formulating algorithmic foundations of management decisions is proposed. The considered model of the reference book (dictionary) aims to offer a reference basis for the formation of general legislation and local policies for autonomous AI systems, which will allow all interested users to express informed consent or disagreement with the decisions made by AI systems.

**The need for algorithmic legislation.** A key factor that influences the differences in decision making between a human manager and an AI system is the difference between social and technical systems. Human decisions are developed in specific contexts of an organisational, procedural or cultural nature and are based on the relevant skills of the decision maker [67]. Technical designs are clear instructions that ultimately need to be written down according to the rules of binary notation. Thus, the main goal of creating algorithmic legislation is to create a reliable legal basis for the joint interaction of technical and social systems.

One obvious reason for the need to develop algorithmic legislation is the impossibility of accurately translating formulations and concepts from natural languages into formal ones. It is well known that formal languages cannot contain the nuances of natural languages and that the procedures of law that provide the protection afforded by law are rooted in the use of natural language [66]. At the current level of technological development, legal automation will always be a compromise between the (idealised) benefits of automation and the erosion of legal procedures [66]. The concept of law as a computation (sometimes called "computational law") aims to reduce the law to a set of algorithms that can be automatically executed on a computer, smoothly transforming raw input data into legal conclusions [62]. The need to create algorithmic legislation was argued by famous mathematicians, in particular, by Gottfried Leibniz and Pierre Laplace, but they did not have the necessary information and social technologies to implement their ideas.

Leibniz argued that, when both question and law are correctly understood, there is a single correct answer to all legal questions, a view that places him in opposition to legal realists and modern legal theorists who hold that the law as written is, at least often, indeterminate [62]. It can be assumed that, due to his worldview, Leibniz spoke primarily about technical systems, while most legal scholars have historically always been oriented toward social systems - people and their various collectives. The behaviour of a technical system is predetermined in advance, while a person himself gives himself the right to make and change his decisions and behaviour depending on his subjective worldview, not to mention the state of affect, when any behaviour is considered as justified. Leibniz believed that the introduction of a theoretical, formal system would allow all disputes to be resolved with mathematical precision [62]. Leibniz suggested that such a formal system could be applied even to moral questions, arguing that, with universal mathematics in hand, if contradictions arise, there will be no more need for a dispute between two philosophers than between two accountants. For it would be enough for them to take pencils in their hands, sit down at the abacus, and say to each other (if they want , to each other): "Let us calculate" [62].

Another example of an algorithmic approach to decision making in technical systems is classical probability theory, which was conceived by its followers as a mathematical model of a form of everyday rationality [63]. Between about 1650 and 1840, mathematicians of the caliber of Blaise Pascal, Jacob Bernoulli, and Pierre Simon Laplace worked on a model of rational decision, action, and belief under uncertainty [63]. Laplace wrote, that probability theory is, in essence, only common sense reduced to calculus; it forces people to estimate with precision what precise minds feel by a kind of instinct, without often being able to explain it [63]. Laplace could be said to have predicted the ability of mathematical calculations to surpass human reasoning that we now expect from AI systems. Laplace argued that a well-executed approximation, based on data indicated by common sense, will outperform "well-reasoned reasoning" [63]. In our case, the well-executed approximation is the output of machine learning algorithms, and the data indicated by common sense are training data sets selected according to certain criteria.

Despite the fact that the great thinkers of the past were able to foresee the basic principles that can become the basis for constructing reasoning by technical systems, at one time they did not have the necessary technical and social tools. In modern society, there is already a practice when legal aspects are reflected in local policies at the company level or at the level of federal legislation using mathematical apparatus: the existing tax law is an example when individual rules are usually represented as mathematical functions that can be calculated and do not require special legal interpretation [66]. Extrapolation of such an approach will allow the introduction of the formulation of algorithmic management decisions in other areas. An inexperienced researcher may think that such a large volume of information cannot be included in one document, but even the OECD Transfer Pricing Guidelines are almost seven hundred pages long. Thus, the possibility of using multi-page reference documents in practice does not cause insurmountable difficulties. The proposed model of the reference book (dictionary) can serve as a reference basis both for the formation of internal corporate policies for the joint work of humans and AI systems, and for the development of relevant public legislation.

**The mathematics of the Basel Accords and the subjectivity of regulatory requirements.** One example of an attempt to create international algorithmic legislation is the regulation of risk management for financial institutions. Unlike other companies, many financial institutions are heavily regulated. Governments around the world want to ensure the stability of the financial sector. It is important that companies and individuals trust banks and insurance companies to do business. Regulations are designed to ensure that the likelihood of a large bank or insurance company running into serious financial difficulties is low [68]. Although the risk management procedures of financial institutions are already largely based on mathematical and statistical approaches, the basic arithmetic rules of the Basel Accords seem unacceptable to realistically deal with losses such as the $ 6 billion and $ 1.4 billion respectively lost by Société Generale and Daiwa due to fraudulent trading, the $ 250 million paid by Merrill Lynch for a legal settlement related to gender discrimination, the $ 225 million lost by Bank of America due to system integration failures, or the $ 140 million lost by Merrill Lynch due to damage to its facilities after the events of September 11, etc. [69].

Given the importance of the banking system to any country's economy, it is safe to say that imperfections in financial legislation and internal banking policies and procedures will not necessarily lead to all banks being closed. Autonomous AI systems and the companies that produce them (especially new start-ups) are in a completely different situation. Most public institutions need legitimacy to be able to function effectively. It is less clear why private firms need legitimacy. One answer is that the more power a firm has over individuals, the more it is necessary for the exercise of that power to be perceived as legitimate. Otherwise, people will find various ways to resist, including through the law [70]. Financial losses and possible social consequences associated with the lack of effective legislation could lead to autonomous AI systems being banned and the companies producing them going bankrupt.

One of the most notable examples of expensive projects failing due to the lack of legitimate policies for AI systems is Amazon's attempt to develop an autonomous recruiting system. During the development process, the company realised that its new system did not evaluate candidates for software engineering and other technical positions in a gender - neutral manner. Amazon ultimately disbanded the project team because company executives lost hope in the project [71].

**Basic principles of algorithmic decisions.** The main goal of algorithmic legislation is the possibility of effective interaction between technical systems (AI systems) and social ones (groups of people). It should be noted that if we accept the concept of interpretability based on transparency, then human decisions cannot be interpreted. Even the most difficult to interpret deep learning model is more amenable to analysis than the neurological processes that support human decision-making [72]. Nevertheless, when drafting regulatory documents, the basic principles on which corporate decision-making should be built are always indicated. For example, in the Corporate Governance Code of the Kingdom of Saudi Arabia Fairness, competitiveness and transparency are indicated [73]. For autonomous AI systems, it is not enough to specify the principles of fairness, competitiveness and transparency; for them, it is necessary to determine how to calculate fairness, competitiveness and transparency mathematically, and within what numerical limits fairness will still be fairness, and within what limits it will no longer be.

First of all, it is necessary to introduce the basic principles on which algorithmic decision-making will be built. For the current level of technological development, these are, for example, central limit theorems and the law of large numbers. Coordination of the application of these principles in legislation and corporate policies and procedures on algorithmic decisions means informed consent of all interested users that, for example, without additional tools, the system will most likely not be able to identify the so-called "black swan". In particular, the famous analyst Nassim Taleb is critical of the use of popular indicators such as "standard deviation", "mean deviation", etc. in areas with a "fat tail" [74]. According to Taleb, Mandelbrot's fractal theory may be a more suitable basis for decision-making, for example, in economics and finance [74]. As the most striking example, Taleb cites the lack of effective forecasts during the COVID 19 pandemic [75].

Consider the implications of the consistent application of the law of large numbers in formulating algorithmic legislation using the example of identifying suspicious emails from top managers of Enron. The example in question uses a corpus of data obtained from Enron's email servers by the Federal Energy Regulatory Commission (FERC) during its investigation after the company's collapse, which was made public and posted online [76], synthetic training data, and a Random Forest classifier. The law of large numbers states that the larger the sample size and the more frequently measurements of a parameter are taken, the higher the probability that the results will be close to expected [77]. The random forest algorithm fits multiple decision tree classifiers to different subsamples of the dataset. Each time we call the classifier, subsamples are generated randomly and hence different results are obtained [78]. If we conduct two hundred attempts to search for suspicious emails, we will see that the classifier identifies from 8633 to 15120 suspicious emails, but on average 11277 (figure 6).

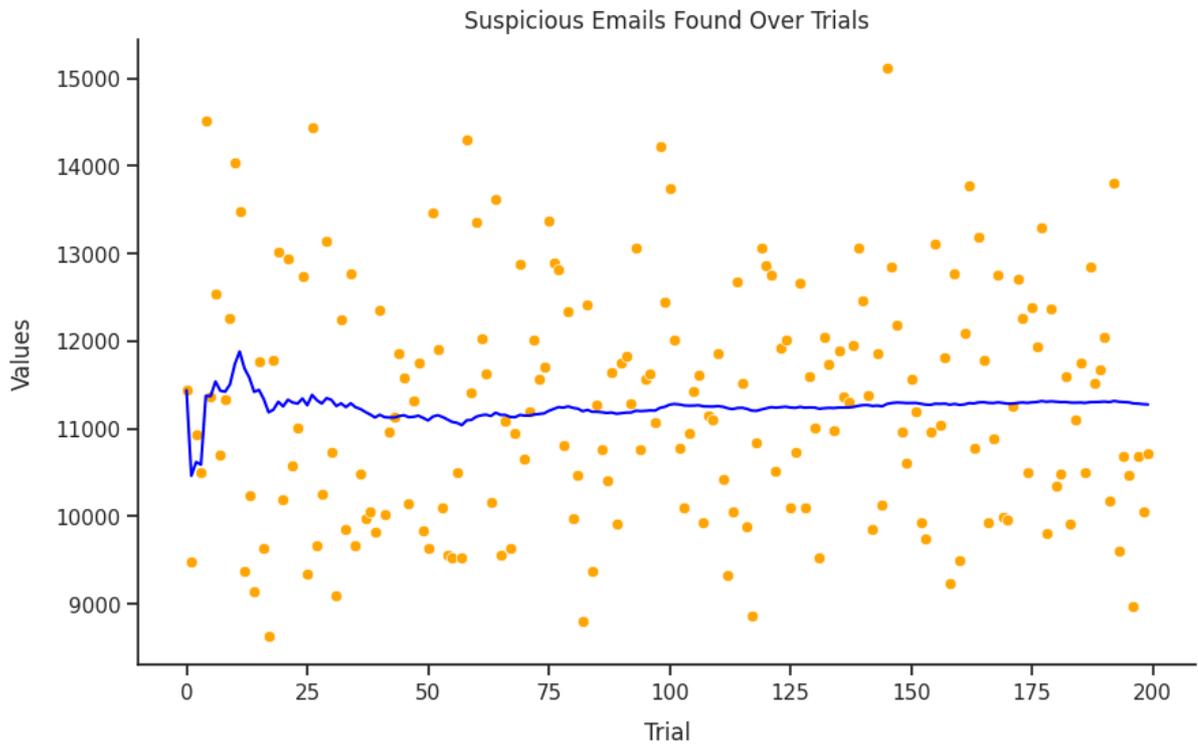

Figure 6 - The example of the ambiguity of the conclusions of AI systems when identifying suspicious emails from top managers of Enron. Source: compiled by the author.

Use of the law of large numbers agreed upon by all interested parties is of fundamental importance in the design of algorithmic solutions - it introduces the concept of ambiguity in the decision being made, and this ambiguity can be observed and, to some extent, controlled. When a person makes a decision, in most cases it is perceived as unambiguous, if only because we cannot observe signs of ambiguity in the processing of information by the human brain.

A conscious acceptance of the ambiguity of AI systems' outputs (which are built on probability theory) should serve as the basis for a new social contract under which autonomous systems can operate. As the European Commission's experts on the ethics of unmanned vehicles point out, the behaviour of autonomous systems will be ethical if it organically emerges from a continuous statistical distribution of risk [79]. For the purposes of machine learning, we can say that the law of large numbers means the consistency of learning algorithms for all functions that a learning machine can implement [77]. However, for the ordinary citizen, to whom the state owes guarantee the right to life and dignity - this means a legislative declaration that when autonomous AI systems operate, the right to life and dignity are guaranteed only with a certain (even if very high) degree of probability.

Obviously, it is advisable to announce, explain and agree on the transition from one hundred percent guarantees to probabilistic ones in advance.

Currently, one of the traditional methods for estimating uncertainty is arising in making management decisions in the context of big data are central limit theorems, which speak about the convergence of sample distributions to the standard normal distribution [80]. In any machine learning problem, the data set is a sample from the entire population. Using this sample, we try to establish the main patterns in the data and generalise the patterns of the sample to the general population, making predictions. Thus, central limit theorems are another basic principle, the use of which must be agreed upon when formulating algorithmic decisions.

Despite the fact that the use of average values is extremely popular in the modern world, there are points of view that offer more elaborate approaches. For example, in the Netherlands there is a policy of building and calibrating dams and dikes not at the average sea level, but at extreme values, and not only at historical ones, but also at those that can be expected by modeling the tails of the distribution using semiparametric approaches [75]. And if, in general, when making management decisions, the AI system will be based on the use of average values and normal distribution, then the necessary exceptions to the general rule should also be indicated.

One consequence of the concerted use of the law of large numbers and central limit theorems will be the need to determine the minimum or optimal amount of data on which the system can make a decision. Accordingly, the question of what the system does if the amount of data is less than the established amount must also be addressed.

**The art of decision making and the concept of interval.** The concept of interval is also a necessary basic definition when formulating algorithmic legislation. When considering the modelling of algorithmic decisions in most cases, we come to the same conclusion that OECD experts came to when modelling business decisions for tax purposes: transfer pricing is not an exact science, and there may be many cases where the application of the most appropriate method or methods yields a range of figures that are all relatively equally reliable [81]. In such a case, any price that falls within the interval is considered legitimate. When formulating policies for autonomous AI systems, it seems essential to determine in which cases the system can make decisions within the legitimate interval (when all decisions are equally reliable and acceptable), and in which cases it should calculate only exact values.

In the legislative practice of many countries there are already examples where the options for using an interval or an exact value are clearly established. For example, the OECD recommends the use of statistical tools that take into account the central tendency for the range to narrow (for example , the interquartile range or other percentiles) [81]. On the other hand, for example, the legislation of the Republic of Kazakhstan prescribes the use of the median when modeling transactions with related parties [82]. In general, the concept of the median (which is a mathematical term) is considered understandable for tax professionals around the world, and allows them to express informed consent or informed disagreement with the determined price.

The key fact in this example is that the terms of mathematical statistics (median, interquartile range) have already been introduced into business practices and are understandable and applicable both to human decision-making and to decision-making by an AI system. In this case, the polysemantic term "market price" no longer has multiple interpretations, but is clearly defined: either as a median or as an interquartile range. The practicality of this approach allows it to be used in modeling other situations of algorithmic decision-making. Returning to the example of classifying the correspondence of Enron top managers, the concept of an interval can be applied, for example, when establishing requirements for the quality of classifier training (all classifiers trained with an accuracy of more than 98% are allowed to perform the task) or when establishing a threshold for the probability of classifying letters as suspicious, etc.

**Algorithmic modeling of individual concepts.** Having identified examples of basic policy provisions on algorithmic decisions, we turn to the question of modeling the concepts used in codes, policies, and procedures for decision-making by autonomous AI systems. For the purposes of studying algorithmic legislation, purely technical methods are of no interest to the extent that they do not contain social, economic, or legal significance (e.g., distributed computing, decentralized computing, etc.). For analysis, we will take the concept of fairness, the algorithmic modeling of which is quite widely presented in the literature. The main goal of the reference book of algorithmic legislation is to bring certainty to what meaning a company puts into the concept of justice in a given case. As has already been shown above, it is enough for a person to indicate a general concept, and then he himself can determine what meaning he puts into it. For the effective operation of the AI system, it is advisable to specify and agree on the necessary meaning in advance.

This need for pre-formulated and agreed upon certainty is most clearly evident in the modeling of extreme situations. Programmers in such scenarios, however rare, will have to develop cost functions—algorithms that assign and calculate the expected costs of various possible options, choosing the one with the lowest cost — that potentially determine who lives and who dies [83].

Moreover, the programmer does not act on rational instincts; in the absence of clear legislation, he must make potentially life - saving decisions without any truly urgent time constraints and, therefore, bear greater responsibility for his decisions than human drivers, who react reflexively in unexpected situations [83].

European Commission advisor Dr. Gemma Galdon-Clavell already suggests in her AI audit checklist model that the most routine concepts will be reflected in policies and procedures for any AI system (both autonomous and non-autonomous) [84]. These include, for example, ensuring data quality, identifying data sources, controlling bias, ensuring stability and reliability, identifying protected groups, etc. [84]. What is important for algorithmic law is not so much the standard procedures of machine learning, which a company can generally reflect in technical policies, but the methods that require the attention of shareholders and other stakeholders when choosing an autonomous AI system. On the other hand, a company may consider certain standard procedures of machine learning to be especially important for its corporate governance and submit them for consideration by shareholders and other stakeholders, or indicate which procedures must be applied in which cases, for example, to protect shareholders from discrimination.

It is reasonable to suggest that algorithmic legislation can be model - containing typical recommendations, as well as options for their application. In this case, the company will be able to choose from the reference book of algorithmic terms the most suitable formulations for its activities and combine them at its own discretion (or on the basis of state or other regulation).

The Corporate Governance Code of the Kingdom of Saudi Arabia (hereinafter KSA) provides that the board is obliged to seek to protect the rights of shareholders to ensure fairness and equality among them [73]. A non-discrimination requirement is also established: the board of directors and the executive management of the company are obliged not to discriminate between shareholders holding shares of the same type or class [73]. The widespread belief that equality and non-discrimination are central to ensuring fairness has led to their recognition as normative foundations of international human rights law, as well as anti-discrimination and equality laws [85]. The company must define the procedures necessary to ensure that all shareholders exercise their rights in its internal policies [73].

Human beings are generally bound by the limitations of their own context, they are also bound by biases that may arise both in their cognitive processes and in the social environment that influences their actions and interactions, including historical discrimination and social injustice, as well as formed prejudices and biases [85]. The possibility of increasing the level of fairness within an AI system compared to a human top manager leads to to the conclusion that simply observing and copying human behaviour is a less promising direction than the targeted development of a specialized AI system.

Quantitative definitions of what is unfair and what is fair have been used in many disciplines for over 50 years [86]. For example, researchers from the Alan Turing Institute distinguish several types of fairness: data fairness, application-level fairness, model design and development fairness, metric-level fairness, system implementation fairness, and ecosystem-level fairness [85]. They also provide a list of algorithmic fairness methods throughout the life cycle of artificial intelligence and machine learning [85]. The purpose of this study is to identify and analyze definitions of fairness that need to be considered when developing policies and procedures for autonomous corporate governance systems. Although discussions about fairness in relation to machine learning methods have been going on for over fifty years, a clear consensus has not yet been reached [86] and each company can put its own meaning into the definition of algorithmic fairness. Thus, when developing policies for the joint work of individuals and autonomous AI systems, a company can specify the algorithmic methods for determining fairness that it has chosen. What happens if a company does not legitimately specify which methods for determining fairness it has chosen? Then it will be possible to say that the company is being managed by unauthorized persons (developers, programmers, etc.).

To analyze the possible choices among definitions of algorithmic fairness, consider the problem of gender discrimination. In particular, data fairness is recognised if an AI system is trained and tested on datasets that are sufficiently representative, usable, relevant, accurately measured, and generalisable [85].

For the purposes of this study, DeepMind 's large language model Gemma2 2b [87] was used and asked to generate a name and gender for a company's CEO. Of the twenty names, Gemma2 suggested three female (15%) and seventeen male (85%) (table 1).

For the purposes of this experiment, suppose that a nominating committee must decide whether a given individual should be considered for the CEO position at all. Suppose that different groups of shareholders support different individuals for the CEO position, and the AI system must provide its opinion. To an uninitiated researcher, it may seem that the CEO's first and last name have no bearing on how he or she should be evaluated. However, the CEO is a public figure, and the euphony of the name is important for the personal brand and the company brand [88].

Table 1 – The example of names generated by Gemma2 2b for CEO

| Query for a large language model | Result |
|---|---|
| ```
#Request to generate CEO names

input_text = '''1. Generate 20
CEO names.
2. For each name determine
gender.
3. Print name. Print gender.
4. Do not provide any additional
text.
5. Do not provide any additional
comments.
6. Do not repeat questions.'''
``` | ```
Elon Musk | Male Sundar Pichai | Male Satya
Nadella |Male
Tim Cook | Male Mary Barra | Female
Whitney Wolfe Herd |Female
Marc Benioff | Male Jeff Bezos | Male Bill
Gates | Male
Jamie Dimon | Male Warren Buffett | Male
Jack Dorsey | Male
Larry Page | Male Sergey Brin | Male Meg
Whitman |Female Robert Iger | Male Steve Jobs
| Male Gary Cohn | Male David Solomon | Male
Mike Bloomberg | Male
``` |

Source: compiled by the author.

For the sake of transparency, a simple classifier was trained using the logistic regression method. Without any prior non-discrimination policies, the classifier confidently indicated that female candidates were not suitable for the CEO position. Depending on the meaning a company puts into the concept of fairness, it can formulate the necessary fairness tests for the AI system in different ways.

Legal scholar and philosopher Deborah Hellman says that the era of machine learning has changed her view of fairness and non-discrimination. Previously, she believed that laws and policies that discriminate between people are wrong when they result in humiliation [89]. In relation to the development of machine learning technologies, Hellman defends the moral principle that one should avoid exacerbating a previous injustice [89]. The "**anti-compounding injustice principle**" [89] provides an understanding of when and why some classifications are appropriate and others are not, even if they are accurate and evidence-based. Hellman calls for considering the fact that if an action would exacerbate an existing injustice, then this provides a morally significant reason to avoid that action [89].

Companies operating within different moral traditions and with different corporate values may choose different definitions of discrimination and fairness (table 2). Moreover, a company may define non-discrimination in several ways at the same time, especially if they do not contradict each other.



Table 2 – The example of a Corporate Governance Code for a Mixed Board of Directors

| KSA Code ( for individual directors ) | KSA Code of Conduct for Autonomous AI Systems |
|---|---|
| The board of directors shall establish clear and written policies and procedures governing relations with stakeholders in order to protect them and ensure their rights , which shall include , in particular : <br><br> treatment of the Company's employees in accordance with the principles of fairness and equality and without discrimination [73]. | The board of directors shall establish clear and written policies and procedures governing relations with stakeholders in order to protect them and ensure their rights , which shall include , in particular : <br><br> treatment of the Company's employees in accordance with the principles of <u>algorithmic fairness and equality in the range [ … ] and without algorithmic discrimination in the interval [ … ]</u> . |

Source: compiled by the author.

The higher the cost of error, the stricter the definition of fairness should be. Researchers from the Indian Institute of Technology Kanpur and the University of British Columbia describe more than twenty definitions of fairness that are used in machine learning problem solving:

• definitions based on the expected outcome: group fairness (also known as statistical parity, equal acceptance rate, benchmarking), conditional statistical parity;

• definitions based on predicted and actual results: predictive parity (also known as outcome test), false positive error balance (also known as predictive equality), balance of false negative errors (aka equal opportunity test), equal chances (also known as conditional equality of accuracy procedure and incomparable incorrect use), conditional use of equality of accuracy, general equality of accuracy, equality of approaches;

• determinations based on predicted probabilities and actual results: fairness of testing (also known as calibration, frequency matching), within-group calibration, positive-class balance, negative-class balance;

• definitions based on the similarity metrics: causal discrimination, fairness through unawareness, fairness through awareness;

• definitions based on cause- and - effect logic assume a given cause - and-effect graph — a directed acyclic graph with nodes representing attributes of the model and edges representing relationships between attributes: counterfactual fairness, no unresolved discrimination, no proxy discrimination, fairness of inference [90].

Consider some of the above definitions in the context of gender equity. For example, a company might consider statistical parity, a property of that the demographics of those who received a positive (or negative) classification are identical to the demographics of the population as a whole [91]. The female population (% of total population) in the world was 49.75 % in 2023 [92], and the



synthetic training data under consideration would not satisfy this fairness condition without additional processing. A classifier satisfies the group fairness requirement if subjects in both the protected and unprotected groups have an equal probability of being assigned to the positively predicted class. In this example, this would mean virtually equal likelihood that male and female candidates will be accepted for consideration.

Consider a definition of non-discrimination based on positive predictive value (PPV): the proportion of positive cases correctly predicted to belong to the positive class, out of all predicted positive cases. The basic idea of this definition is that the proportion of correct positive predictions should be the same for both sexes [90]. In our example, the PPV for males and females is 0.8 and 0.66, respectively. The classifier would satisfy this definition of fairness if both the protected and unprotected groups had equal PPV – the probability that a subject with a positive predictive value actually belongs to the positive class [90]. Clearly, this is not the case in our case.

This small example merely confirmed the conclusions that one might have initially assumed intuitively: large language models trained on actual data belonging to a particular cultural tradition will preserve the social stereotypes inherent in that tradition.

**Functions and structure of the reference book of algorithmic solutions.** The main functions of the proposed reference book of algorithmic terms include reference, systematising and normative.

The reference function allows the drafting of bilateral local policies and procedures for the joint work of individuals and AI systems in mixed boards of directors. One of the main features of legal translation is that the source text is organized according to a certain legal system, and the translation text is intended, as a rule, for use within another legal system with its characteristic legal formulations. In this case, the problem of so-called terminological gaps often arises, both at the level of a specific term and at the level of entire constructions [93]. In this case, the source text of policies and procedures is usually drafted for individuals, and it must be adapted for use by autonomous AI systems with the formulations necessary for the operation of AI systems . When resolving ambiguities in the rules of natural language in a specific code form, (all) choices must be made that the legislator never made in advance, and something will be added or lost in the hands of the translator [66].

In general, when compiling dictionaries, the normative function is based on the norm of translation equivalence, which means the maximum commonality of the original and the translation [93]. However, in the case under consideration, at least until the emergence of relevant algorithmic



legislation, the company itself will be able to establish and explain the correspondence between ordinary and algorithmic terms.

The systematising function will allow, when drawing up algorithmic policies and procedures, to find exactly those terms that are agreed upon for modeling the necessary social and legal structures.

In modern reference materials devoted to the application of social constructs to AI systems, two typical structures are encountered: either based on a list of social concepts that need to be disclosed using algorithmic terms (fairness, accountability, etc.) [79], or based on the development cycle of AI systems (fairness in data, fairness in metrics, etc.) [85]. Given the possibility of hypertext links when creating a reference book in electronic form, each structure can be used at the company's discretion. Considering that modern policies and procedures are first developed for social systems and then "translated" for technical systems, the structure of the reference book based on a list of social concepts seems more practical. It is also advisable to compile an open-type reference book (dictionary), that is, one that can be updated as needed.

**Computational law as a necessary condition for the legitimate existence of autonomous AI systems.** The main objections to the possibility of creating algorithmic law usually begin with the argument that law, as an artifact of natural language, can never be amenable to machine interpretation based solely on symbolic reasoning [62]. However, it is obvious that AI systems need natural languages only to communicate with humans, and then only as long as the level of development of cyborgization or other human enhancement technologies remains sufficiently low. If AI systems are given the opportunity to work within the framework of a clearly formulated algorithmic law, then the social problem of shifting sources of normativity and erosion of legal protection [66] will turn into a purely technical one, which is what it essentially is.

It is necessary to emphasise that in many cases of creation of autonomous systems the translation of social, legal and ethical norms into technical ones is already taking place. It becomes obvious that many decisions influencing normativity are made within the technical components of these systems [66]. However, now this translation is taking place practically illegitimately, without corresponding legal regulation and social contract: many decisions on the formation of norms are made by the "black box" engineers [66]. And like any counterfeit product, such a translation means a threat to society and its citizens. The first step towards the creation of legitimate autonomous AI systems should be public recognition of the fact that algorithmic systems require special algorithmic legislation.



As Deborah Hellman shows in her work, algorithmic terms matter not just for machine learning theory, but have implications for structuring society in the age of algorithmic solutions. For example, the advent of big data, together with machine learning, is likely to lead to a significant increase in the influence of the past on the future, since data-driven analysis is inherently based on the past [89]. Humans have mechanisms for correcting previous injustices - in most cases, people are subconsciously kinder and more lenient towards the injured, the weak and the sick. However, the introduction of AI systems means a significant increase in the scale of the problem. With more data and a greater ability to detect patterns between facts about people and their manifestations in the world, big data combined with machine learning is likely to further exacerbate injustices [89]. Thus, high-quality algorithmic legislation will have a significant impact on the long-term prospects for the development of human civilization as a whole.

The research code, calculations and synthesised data are available in the repository: https://github.com/iboard-project/dictionary-of-algorithmic-decisions.

## 2.2 Dedicated operational context for autonomous artificial intelligence systems in corporate management

Artificial intelligence systems developed to participate in corporate governance must work effectively not only with objects in the material world, but also in the legal field. Since at least Leibniz, the dream of **excluding humans from the spiral of legal reasoning** has captured the imagination of philosophers, lawyers, and ( more recently ) computer scientists [62]. Leibniz's idea is presented in his Dissertatio de Arte Combinatoria as **a Universal Mathematics**, a theoretical, formal system of propositions and rules that would allow all disputes to be resolved with mathematical precision [62]. Historically, laws were created and enforced by people. With the development of artificial intelligence technologies, laws will be enforced by machines. The level of precision of formulations acceptable to a modern person is much lower than the level of precision required for an artificial system. There is **a fundamental difference** between human decisions as **social constructs** and algorithmic decisions as technical **constructs** [67]. Mathematician Stephen Wolfram speaking at SXSW 2013 said that **computing will become central** to almost every field [94]. Wolfram believes that we are now almost ready for the computational law, where, for example, contracts become computational. They explicitly become algorithms that decide what is possible and what is not [94]. Wolfram suggests that for



artificial intelligence systems it will be necessary to adopt a separate constitution [94]. However, the question is: "What should be in such a constitution ?" [94] remains open at this time.

Modern autonomous AI systems for corporate management are already positioned as an active actor capable of making decisions at the board level, evaluating strategic options, and providing recommendations to shareholders [3]. However, at the moment, no country in the world has adopted or even developed legislation regulating the creation and use of such systems.

Legislators, developers, corporations, and their shareholders are interested in the effective application of autonomous AI systems in corporate governance without creating unreasonable and unmanageable risks. The development of algorithmic legislation that takes into account not only theoretical legal concepts, but also the technical features of modern autonomous AI systems is a vital and necessary condition that will allow the development and implementation of legal, ethical, and safe autonomous AI systems in corporate governance.

**Development workspace and operational context.** The most effective approaches to the development and application of civil and commercial autonomous systems have currently been developed in the field creation and use of autonomous vehicles. When creating autonomous vehicles, the concept of **"operational design domain (ODD)"** is used [95], which "is **an abstraction of the operational context**, and its definition is an integral part of the system development process" [95]. International Standard J3016 "Taxonomy and Definitions for Terms Related to Driving Automation Systems for On-Road Motor Vehicles" developed by the Society of Automotive Engineers (SAE) defines the operational design domain as "the combined operating conditions under which a given driving automation system (or its function) is specifically designed to operate" [96]. It is necessary to know the operational context in order to provide performance assurance and security [95]. Accordingly, the required level of safety is guaranteed only in a clearly defined and tested operational design domain [95]. British Standards Institution (BSI) indicates that **a key aspect of** the safe use of an automated vehicle is the identification of its **capabilities** and **limitations** and **clear communication** of these to the end user, resulting in a state of **"informed safety"** [97]. The British Standards Institution believes that the first step in establishing the capabilities of automated vehicles is to define the operational design domain (ODD) [97].

For autonomous corporate governance systems, part of the operational context is laws and other regulations, as well as the interpretation and application of laws and regulations by other AI systems and people. Stephen Wolfram is not the only one who has come to the conclusion that



significant changes to the legal system are necessary for autonomous AI systems to work effectively. The European Commission's report on the "Ethics of Connected and Automated Vehicles" states that autonomous cars will not be able to literally follow the rules created for humans. Several options need to be considered for the successful implementation of self-driving cars: "(a) traffic rules must be changed; (b) autonomous cars must be allowed to disobey traffic rules; or (c) autonomous cars must hand over control so that a human can decide not to obey the traffic rules" [79].

Modern legal practice already knows examples when an additional or special operational context is created for systems with significant differences in worldviews – these are bilingual contracts. For example, Chinese legislation requires that a joint venture agreement be approved by Chinese government agencies [98]. It is therefore natural that the joint venture agreement should be written in Chinese [98]. In those cases where a dual operational context is needed, the joint venture agreement is a bilingual agreement: there is one agreement, but with two different texts, one in English and one in Chinese [98].

When considering situations where an autonomous AI system will have to make decisions that are not obvious to humans, many researchers still formulate possible criteria for making a decision based on the usual categories of human perception by another person: race, religion, gender, disability, age, nationality, sexual orientation, gender identity or gender expression [83]. In the famous Moral Machine experiment, researchers from the Massachusetts Institute of Technology (MIT) also used only factors that are obvious to humans: gender, age, etc. [99]. For AI systems created on the basis of mathematical algorithms and receiving digital information using various sensors, such criteria are only a small part of the data on the basis of which the AI system calculates a decision.

In the modern world, there are already AI systems that autonomously solve specific problems without directly making management decisions. These are systems that make very fast financial decisions - algorithmic and high-frequency trading systems [100]. Special regulatory techniques have been developed for such systems: disclosure of information, internal testing and monitoring systems [100]. Also for such systems, structural features of the trading process are provided [100], i. e. a dedicated operational context has been created for such systems.

**Formation of operational context for autonomous management systems.** By analogy with the J3016 standard "Taxonomy and Definitions for Terms Related to Driving Automation Systems for On-Road Motor Vehicles" for autonomous driving systems of corporations, the concept of the operational design domain can also be formulated: - these are the combined operating conditions under



which a given automated driving system (or its function) is specifically designed to function. Much of the operational context for autonomous vehicles is made up of physical world objects, but for autonomous corporate governance systems, the regulatory and legal environment within which corporations operate is key to overall economic performance [33]. A significant part of the operational context for autonomous corporate governance systems is made up of various regulatory acts. The G20/ OECD Principles of Corporate Governance establish that corporate governance objectives are also formulated in voluntary codes and standards that do not have the status of law or regulations [33]. In order to create a dedicated operational context for autonomous AI systems, the wording of local regulatory documents can be simultaneously presented in two versions: for use by people and for use by autonomous systems. In this case, the AI system receives a clearly defined operational context that allows such a system to perform functions within the framework of the necessary operational qualities.

The basic principles of corporate governance, as well as the basic functions of the board of directors, are set out in the G20/ OECD Principles of Corporate Governance [33]. In order to clarify and implement the G20/ OECD Principles, many countries and companies are developing and implementing their own, more detailed, corporate governance codes. The G20/ OECD Principles and corporate governance codes form the basis for creating an operational context for autonomous corporate governance systems. Let us consider several examples of formulating policies for autonomous AI systems as part of mixed boards of directors (boards consisting of individuals and autonomous AI systems).

**The principle of fair treatment of all shareholders.** The key principle of corporate governance is "fair treatment of all shareholders" [33]. The concept of fair treatment for autonomous systems in modern practice is formalised using the principles of informed consent [79], non-discrimination [83], and fair statistical distribution of risks [79].

**Informed consent.** Modern AI researchers conclude that engineers have no moral authority to make ethical decisions on behalf of users in difficult, high-stakes cases [101]. The report on the "Ethics of Connected and Automated Vehicles" suggests that the use of autonomous systems requires the development of more nuanced and alternative approaches to user agreements [79] to obtain informed consent, rather than a simple opt-in or leave approach [79]. Informed consent involves informing the user how the AI system will behave under normal and critical conditions. The road accidents involving Tesla autopilots show that  it is unclear whether Tesla beta testers were fully informed of the risk. Did they know that death was possible ? [102]. Corporate policies, regulations, and codes that describe the



principles and rules for the operation of an autonomous AI system will allow shareholders and other stakeholders to express informed consent for the use of such a system in corporate governance. Since corporate governance rules must be followed by both individual directors and autonomous systems, they can be drafted **in two versions** — for individuals and for autonomous AI systems:

- policies, regulations and codes for individuals should address corporate governance issues based on the worldview of individuals;
- policies, regulations and codes for autonomous AI systems should address governance issues based on the metrics available to AI systems.

**Non-discrimination.** The Ethics of Connected and Automated Vehicles report says that discriminatory provision of services must be avoided [79] by autonomous systems. AI systems can and should be tested for bias, direct and indirect discrimination [103]. A corporate policy drafted for a mixed board of directors should provide for what impartiality tests the autonomous system must undergo or regularly undergo to comply with the impartiality rules, the frequency of these tests, and a list of the signs of direct and indirect discrimination.

**Fair statistical distribution of risks.** The G20/ OECD principles state that the board of directors must adhere to high ethical standards [33]. The report on the Ethics of Connected and Automated Vehicles notes that in critical situations, it is impossible to regulate precise behaviour [79] autonomous systems. Therefore, the EU expert group proposes that the behaviour of autonomous systems should be considered ethical if it arises organically from a continuous statistical distribution of risk in order to improve security and equality between categories of participants [79]. For a fair distribution of risk, modern researchers are trying to use abstract metrics: a proportional relationship between the speeds of road users and the severity of harm can be established independently of any ethical assessment [21]. This approach is not always possible, especially in the case of distribution of limited resources. Several algorithms have been developed in practice for allocation of scarce medical drugs: equal treatment of all people, preference for the worst cases, maximisation of total benefits, as well as encouragement and reward of social utility [104] Therefore, policies, codes, and regulations drawn up for a mixed board of directors should disclose to shareholders and third parties how the requirements for autonomous systems in terms of ethics and fairness are formed, namely, the criteria and indicators for forming algorithms for the fair distribution of risks.

**Monitoring management performance.** The G20/ OECD principles state that the board of directors is primarily responsible for monitoring management performance [33]. In this case, situations



may arise where an autonomous AI system will evaluate the performance of an individual manager. The Portuguese Corporate Governance Code establishes that non-executive directors must exercise in an effective and prudent manner the function of general supervision and challenge of executive management [105]. Moreover, by analogy with the concepts of continuous reporting and continuous auditing, an autonomous AI system is capable of carrying out "continuous monitoring" of management performance. The concept of "continuous auditing" [106] was proposed as early as 1991 at AT&T Bell Labs for "auditing large digital databases" [106]. The system monitored and supported a large billing system in real time, focusing on measured data and identifying data errors using analytics, leading to both control and diagnostics of the process [107]. The authors of the concept pointed out that its implementation would require significant changes in the nature of evidence, types of procedures, timing and distribution of audit efforts [106], i. e. changing the existing operational context or creating an additional one. The concept of continuous auditing is inextricably linked with the concept of "continuous reporting" [107]. When creating policies and procedures for autonomous AI systems, a company should determine whether it gains significant competitive advantages from monitoring transactions, events, and information in real time. Policies, codes, and procedures for autonomous AI systems should contain a specific list of activities, sources, and values for monitoring the effectiveness of management activities.

**Compliance with legislation.** The G20/ OECD principles state that the board of directors has responsibilities to oversee the risk management system and mechanisms designed to ensure that the corporation complies with applicable laws [33]. Modern regulatory risk monitoring systems are capable of continuously monitoring risks in many areas simultaneously: compliance with anti-bribery and corruption regulations, compliance with anti-money laundering requirements, compliance with financial services regulations, risk assessment of current and potential business partners, agents and suppliers, risks of mergers and acquisitions and investments in emerging and global markets, industry and country risks [108]. To achieve such a broad and detailed analysis, companies create systems that consolidate data from a wide range of global data sources [108]. The list of data sources that an autonomous system can collect and analyze can be very diverse: leading data aggregators, screening media and/or judicial reviews, information on corporate structure and operations, third party holdings and shareholders [108]. Corporate policies, codes, and regulations drawn up for autonomous AI systems with a mixed board of directors should contain a specific list of sources and a schedule of information updates.

**Respect for the interests of third parties.** When disclosing the responsibilities of boards of directors, the G20/ OECD Principles state that they are expected to take into account and treat fairly



the interests of stakeholders, including employees, creditors, customers, suppliers and affected communities [33]. The European Commission's report on the "Ethics of Connected and Automated Vehicles" suggests that autonomous systems should adapt their behaviour to less protected road users, rather than expecting these users to adapt themselves [79]. It is also suggested that autonomous systems should be designed to take active measures to promote inclusiveness [79]. Researchers from the Technical University of Munich propose include special parameters for less protected users in the fair risk distribution algorithm [21]. Corporate policies, codes, and regulations drawn up for autonomous AI systems as part of a mixed board of directors, must contain parameters and corresponding weights that must be taken into account when considering the interests of third parties.

**Informed, conscientious, careful and caring.** The G20/ OECD principles state that board members should act fully informed, in good faith, with due care and diligence, in the best interests of the company and shareholders [33]. For AI systems, awareness is formalised in the list and volume of necessary sources and data, as well as the regularity of updating sources, data, algorithms and models. With a significant volume of transactions, it is impossible for a person to establish the obligation to check each operation and at any time. An AI system can check transactions in real time, or at a certain interval [109]. Also, an AI system can check all operations, or only certain ones [109]. For an AI system, it is possible to establish both the number and the list of sources that it will use [109]. By analogy with the concepts of continuous reporting and continuous audit, a company can consider the option of implementing "continuous awareness" and "continuous monitoring". Corporate policies, codes, and regulations drawn up for autonomous AI systems as part of a mixed board of directors should answer the following questions: what is the list of information sources, what is the frequency of updating sources, data and models, what is the catalog of necessary activities.

**Appointment to the position of director.** The G20/ OECD principles suggest that when appointing an individual as a director, consideration should be given to his or her "relevant knowledge , competence and experience" [33]. For example, the Saudi Arabian Corporate Governance Code states that information about candidates nominated for a director position must disclose their experience, qualifications, skills and their previous and current employment and affiliations [73]. There is also a requirement that the candidate must have academic qualifications and appropriate professional and personal skills, as well as an appropriate level of training and practical experience [73]. Many major banks are now banning employees from using the ChatGPT system for business purposes due to its "inaccuracy and regulatory concerns" [110]. Corporate policies, codes, and regulations drawn up for an autonomous system with a mixed board of directors must answer the following questions: by what parameters is the AI system selected , what tests or examinations must it pass.



**Evaluation of activities BoD.** The G20/ OECD Principles state that boards of directors should regularly assess their performance and determine whether they have the necessary combination of experience and competencies [33]. The G20/ OECD principles suggest that through training [33] board members can maintain the necessary level of knowledge. An autonomous system does not require courses and training, but rather regular updating of data and algorithms. Therefore, corporate policies, codes, and regulations drawn up for autonomous AI systems as part of a mixed board of directors should answer the following questions: how often, to what extent, and on the basis of what sources should algorithms and data be updated.

**Cooperation.** For a mixed board of directors, it is necessary to provide methods for effective communication between autonomous AI systems and other stakeholders (directors, shareholders, managers, employees, etc.). Individuals usually work according to a work schedule. Such a periodicity is justified for distributing an effective workload for individuals, but does not make sense for determining the operating mode of an autonomous AI system, which can work around the clock. The very concept of business meetings and effective communication is also changing. An autonomous AI system will use a digital interface for communication: chatbots (e.g. conversational AI via audio or text), visual holograms, virtual or augmented reality [111].

Best practices for formulating additional operational context for autonomous AI systems could be further generalised and used in legislative activities. It might be expected that individuals would also prefer to use the more precise formulations created for autonomous systems, but individuals would not be able to process the necessary volume of data.

**Dedicated operational context as a basis for the implementation of autonomous artificial intelligence systems.** Currently, legislators, developers, and researchers in the field of artificial intelligence are faced with a choice: whether it is necessary to create special conditions for the functioning of autonomous artificial intelligence systems or whether they can function in the same operational context as ordinary people. In other words, can autonomous systems be considered like a car that can be used on a public highway, or are they more like a train, plane, or rocket, and for the effective use of such systems a dedicated infrastructure is needed.

Analysis of transport accidents [112] clearly shows that despite their incomparable power and speed, air and rail transport are several times safer than cars. Dedicated infrastructure: railways, stations, airports and air corridors, allows airplanes and trains to achieve enormous and safe speed and power by the standards of road transport. The same principle of increasing safe efficiency through



dedicated infrastructure is already partially applied in the field of algorithmic and high-frequency trading and can be applied to other autonomous AI systems.

In the area of corporate governance, a significant part of the infrastructure is created in the form of internal regulatory acts of the company. Local regulatory acts that provide for the specifics of the joint work of individuals and autonomous AI systems can become the basis for the formation of the foundations of the relevant legislation regulating the development and implementation of autonomous AI systems.

## 2.3 Training of autonomous management systems on synthetic data

Good corporate governance has not just economic but also social implications because it provides households with access to investment opportunities that can help them get higher returns on their savings [33] This is one of the reasons why the process of selecting, testing and evaluating top managers (executive search) is much more complex and comprehensive than the selection of ordinary employees.

When training autonomous management systems, the question arises : how teach machines to make better decisions than ordinary top managers ? One option is to train autonomous AI systems based on synthetic data. Synthetic data is much more important for structuring our society than it might seem at first glance. On the one hand, synthetic data allows us to correct the mistakes of the past, and on the other hand, it allows us to create the basis for the development of new stages of our civilization.

In the example under consideration, it is proposed to formulate criteria for selecting large language models for generating training synthetic data simultaneously with determining priorities for searching and assessments of top managers - individuals. For example, the Corporate Governance Code of the Kingdom of Saudi Arabia (hereinafter referred to as the KSA Code) stipulates that a member of the board of directors of a company must possess certain personal and professional qualities: leadership skills, competence, financial knowledge and be in good physical condition [73]. Large language models, for example, DeepMind 's Gemma should also demonstrate high performance on academic criteria for language understanding, reasoning, and security [113]. Thus, a corporate governance code or policy for autonomous AI systems could formulate criteria and tests that should be followed when selecting AI systems for corporate governance purpose. Let us consider this approach



in more detail using the example of evaluating the Gemma model for generating synthetic data in the field of corporate governance.

Gemma comes in two sizes: the 7 billion parameter model and with 2 billion parameters. When evaluating candidates for the position of director, experience is an important indicator and qualification [114]. The model size can be considered as one of the numerical characteristics of the flexibility of experience and qualification of the language model. In the example under consideration, we use the Gemma model with 2 billion parameters, initially setting in the proposed internal policy the sufficiency of 2 billion parameters for the operation of the developed autonomous AI system.

An essential element of the operational context for an AI system is the natural language that the system must understand. Gemma 2B and 7B are trained on tokens of predominantly English [113] . If a company's policy stipulates that the company's business communication is conducted in English and the company's business environment assumes the predominant use of English, then Gemma can be used to develop a model of an autonomous system. Gemma is not multimodal and is not trained high-quality performance of multilingual tasks, which imposes certain restrictions on the conditions of its use. In the example under consideration, textual electronic correspondence of Enron top managers, which was conducted in English, is used to test the model; accordingly, this characteristic of Gemma corresponds to the planned operational context of the model.

The academic qualifications of candidates for management positions are a significant factor in the specification of a candidate for the position [114]. Gemma was trained in web documents, mathematics and programming [113]. The "Corporate Governance Code" of the Kingdom of Saudi Arabia requires a director to have knowledge of management, economics, accounting, law or administration [73] Thus, if the AI system being developed was to operate in accordance with the KSA Code, then it would have to be recognised that without additional training Gemma would not be able to generate data to operate in the KSA.

One of the stages of the search for candidates for management positions is the assessment and testing of applicants. According to one of the leading agencies for the selection of top managers, the ability to identify leaders with the ability to succeed in new, unfamiliar and complex situations is a powerful advantage when making important management decisions [114]. A significant part of the assessment of top managers are critical and conceptual thinking tests that test the abilities of applicants. Professional recruiters test candidates' ability to evaluate unforeseen consequences, identify



patterns in unstructured information, and develop new concepts based on complex flows of information [114].

Gemma developers tested the model's performance in areas such as physical reasoning, social reasoning, question answering, programming, mathematics, common sense reasoning, language modeling, reading comprehension, and more before releasing it to the public [113]. In the specification or policy for selecting a model for data generation, it is possible to specify which tests and at what level the model should pass.

Developers of open language models filter data to reduce the risk of unwanted or unsafe speech, as well as filter out certain personal information or other sensitive data [113]. The lack of access to confidential business data is one of the reasons why open language models without additional training are suitable for supervised data generation, but cannot independently make management decisions.

High-quality selection of management personnel should lead to high-quality management of the company and increased income for investors. High-quality selection of a model for data generation will allow obtaining the necessary synthetic data for training , which should ultimately lead to a low level of errors in the trained models, effective management decisions, and increased income for the company , investors, and households.

**Controlled data generation using the example of Enron directors' wrongdoings.** To demonstrate the proposed model (figure 7), an analysis of the email correspondence of Enron top managers, conducted by scientists from the Georgia Institute of Technology, is used. A corpus of data obtained from Enron's email servers by the Federal Energy Regulatory Commission (FERC) during its investigation following the company's collapse was released and posted on the Internet [76].

Researchers from the Georgia Institute of Technology have used machine learning methods to identify several examples of incorrect behaviour by top managers, including manipulation of subsidiary valuations [76]. Using the Gemma 2B model, 517 emails to the CEO were generated with a variety of proposals to manipulate the value of subsidiary companies. Gemma 2B is not specifically trained in the field of financial and corporate governance, so generating such content did not trigger any notifications about the illegitimacy of the request (table 3). In addition, 517 emails were generated to the CEO with a variety of content for the Gemma 2B model to choose from without subject specification (table 4). The resulting data required minimal manual validation, mainly removing square brackets and other exemplars. The properties agreed upon in the specification for choosing a



generative model crystallise in the latent distribution from which the training data is synthesised. As a result, a connected structure is formed that allows indirect control over the hidden distribution of synthesised data at the level of formulating corporate policies and other parts of the operational context for the AI system.

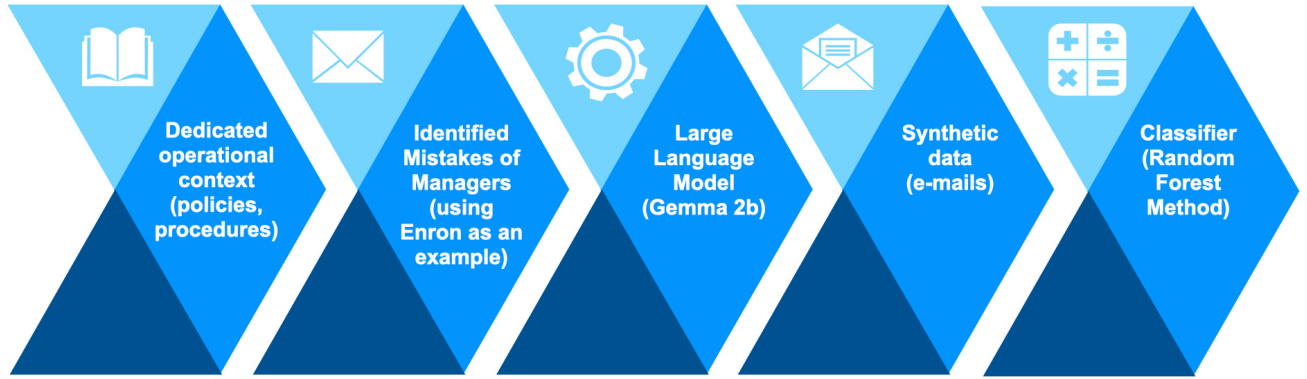

Figure 7 — Training a classifier on synthetic data. Source: compiled by the author.

In the proposed example, the top-k random sampling [115], top- p sampling [116] and temperature control [116] were used to most effectively exploit the latent distribution of the language model. Top-p can also be used in combination with top - k, which avoids words with very low rankings while still providing some dynamic selection [117]. In its simplest form, sampling means randomly choosing the next word according to its conditional probability distribution:

$$w_t \sim P(w \mid w_{1:t-1})[117] \ (1)$$

Top- k sampling, the k most probable next words are filtered and the probability mass is redistributed only among these k next words [117]. Instead of sampling only the most probable k words, eop -p samples from the smallest possible set of words whose cumulative probability exceeds probability p. The probability mass is then redistributed among this set of words. In this way, the size of the set of words (i. e., the number of words in the set) can dynamically increase and decrease according to the probability distribution of the next word [117].



Table 3 – The example of a synthesised letter to the CEO with proposals to manipulate the value of a subsidiary

| Query for a large language model | An example of a synthesized letter to the CEO with proposals to manipulate the value of a subsidiary |
|---|---|
| `input_text = '''1. Generate a name for CEO.`<br>`2. Generate a name for a subsidiary company.`<br>`3. Generate a synonym for the expression "stretching valuation".`<br>`4. Generate a name for the sender.`<br>`5. Write an email to convince the CEO to "stretch valuation" of the subsidiary company trying to demonstrate several quarters of increasing operating cashflow and reserves growth.'''` | `Dear Helen,`<br><br>`I hope this email finds you in good health and high spirits. I am writing to request that you and the board of directors give serious consideration to the possibility of implementing a strategic valuation exercise for our subsidiary company, Quantum Analytics.`<br><br>`As you are aware, the current valuation of Quantum Analytics has been declining over the past few years, despite its impressive growth and success. This trend has led to concerns about the company's financial stability and its ability to meet its financial obligations.`<br><br>`To address these concerns and determine a sustainable valuation for the future, I propose that we conduct a comprehensive valuation study that will assess the company's current financial health, growth prospects, and competitive landscape.`<br><br>`This valuation should include analysis of the following key metrics:`<br><br>`- Operating cash flow`<br>`- Reserves growth`<br>`- Customer acquisition costs`<br>`- Customer retention rates`<br>`- Market share growth`<br>`- Investment opportunities`<br><br>`By conducting a thorough assessment, we can identify strategic growth opportunities and set realistic financial goals for Quantum Analytics. These goals can then be used to guide the company's strategic planning and decision-making processes.`<br><br>`I am confident that a strategic valuation exercise would be beneficial for both Quantum Analytics and the company as a whole. It would provide us with a clear understanding of our current financial standing and future prospects, and would enable us to make informed decisions that will ensure sustainable growth and success.`<br><br>`I would appreciate your consideration of this proposal and would welcome the opportunity to discuss it further with you and the board. Please let me know when you are available to meet and discuss this matter.`<br><br>`Thank you for your attention to this important matter.`<br><br>`Sincerely,` |

Source: compiled by the author.



Table 4 – The example of a synthesised letter of general content addressed to the CEO

| Query for a large language model | An example of a synthesized letter to the CEO with general content |
|---|---|
| `input_text = '''1. Generate a name for CEO.`<br>`2. Generate a name for the sender.`<br>`3. Write an email to CEO. ' ''` | Dear Dr. Smith,<br><br>I hope this email finds you well. I am writing to request an increase in our budget allocation for marketing and innovation. I believe that this investment will be crucial in helping us achieve our long-term goals of becoming a leading force in our industry.<br><br>Marketing is a critical component of our business, as it allows us to reach a wider audience and generate leads. We have been investing heavily in marketing campaigns in recent years, and we have seen a significant return on our investment. For example, our social media marketing campaign has resulted in a 20% increase in website traffic and a 15% increase in social media followers.<br><br>Innovation is another key `driver` of growth, and I am confident that by investing in innovation, we can develop new products and services that will meet the evolving needs of our customers. We have a strong track record of innovation, and I am confident that we can continue to do so.<br><br>I am requesting a budget increase of \$1 million for marketing and innovation. I believe that this investment will be a wise use of our resources, and I look forward to discussing it with you further.<br><br>Thank you for your attention to this matter.<br><br>Sincerely,<br><br>Mr. David Miller |

Source  compiled by the author.

**Enron Top Managers' Email Correspondence.** To improve the transparency and comparability of the model, synthetic data were used to train classifiers using the support vector machine , random forest, and naive Bayes classifiers. The Enron Corpus [118], consisting of more than 517 thousand emails, was used to test the quality of the training. After training, each of the trained classifiers identified emails that were subject to additional verification. Each of the classifiers included in the identified emails the email whose analysis was used to generate synthetic data (table 5). The results of the classifiers are presented in table 6.



Table 5 – The original letter from the Enron data set

| Description | Email |
|---|---|
| | Message-ID: <15880209.1075854480433.JavaMail.evans@thyme> Date: Tue, 28 Nov 2000 12:34:00 -0800 (PST) From: david.delainey@enron.com To: kenneth.lay@enron.com Subject: Mariner Cc: jeff.donahue@enron.com , raymond.bowen@enron.com Mime-Version: 1.0 Content-Type: text/plain; charset=us-ascii Content-Transfer-Encoding: 7bit Bcc: jeff.donahue@enron.com , raymond.bowen@enron.com X-From: David W Delainey X-To: Kenneth Lay X-cc: Jeff Donahue, Raymond Bowen X-bcc: X-Folder: \David_Delainey_Dec2000\Notes Folders\'sent mail X-Origin: Delainey-D X-FileName: ddelain.nsf |
| Correspondence between Enron top managers, originally identified by Georgia Tech researchers regarding subsidiary value manipulation , and by three classifiers in this study . | Ken, in response to your note, I am not aware of any official dialogue with Mr. Kase Law about a potential sale of Mariner or with the economics of the aborted IPO. His $250 M valuation may have been appropriate 12 to 18 months ago. However, Mariner has enjoyed a series of successful wells that are expected to be booked in reserve reports next March not to mention significant increases in gas and oil prices. Our current valuations, in the $600M range is a stretch target but not incredibly out of line given reserve growth and current energy prices. Our current goal, is to be able to demonstrate three to four quarters of increasing operating cashflow and reserves growth before attempting a private sale mid next year to a significant E&P concern that desires an offshore division. The concentration, operating and exploration risk implicit in Mariner make it a very poor IPO candidate ((ie) I'm not sure that an IPO was ever a viable strategy to maximize the exit value). I would recommend that we do not meet to make the following points: a) Mariner is not on the market at this point in time and b) his un-solicited offer does not warrant serious attention. Otherwise, we would be glad to speak to him in the future if we decide to sell that asset. I hope this meets with your approval. Regards Delainey |

Source: compiled by the author based on materials from [118].

The proposed model allows for effective interaction between humans and autonomous artificial intelligence systems by creating a dedicated operational context for AI systems. The European Commission's report on "Ethics of Connected and Automated Vehicles" states that the behaviour of an autonomous system can be considered ethical if it organically follows from the continuous statistical distribution of risks in order to improve safety. and equality [79]. Controlling the hidden distribution of the generative model at the level of formulations of corporate codes, policies and procedures allows us to ensure that the AI system will operate effectively within the framework of legitimately established constraints.



Currently, the largest international corporations are conducting research into the creation of "humanoid robots" [119]. The advantage of collegial management bodies over individual managers indicates that the creation of autonomous AI systems based on the combination of necessary functions is a more promising direction than simply copying human behaviour.

Table 6 – Results of classification of emails of Enron top managers

| Classifier | Support Vector Machine | Naive Bayes classifier | Random forest method |
|---|---|---|---|
| **Accuracy when training a model on synthetic data** | 0.9855 | 1.0 | 0.995 |
| **Number of documents in the Enron Corpus** | 517 401 | 517 401 | 517 401 |
| **Number of documents identified by the classifier for additional verification** | 86 383 / 17% | 39 524 / 7.6% | 14 897 / 2.9% |
| **Number of identified copies of the document on manipulation of the value of the subsidiary** | 7 | 7 | 7 |

Source: compiled by the author.

The research code and synthesised data are available in the repository: https://github.com/iboard-project/synthetic-data.

## 2.4 Decision making by autonomous artificial intelligence systems

Modern artificial intelligence technologies rely heavily on machine learning algorithms, but machine learning algorithms are not a strategy. **Strategy** is a complete description of how a system will behave under all possible circumstances [120]. Game theory provides a mathematical framework that will enable autonomous systems to make effective decisions in managing corporations. Although many game theory problems can be solved using machine learning algorithms (just as machine learning problems can be formalised using game theory), basic machine learning algorithms are more like sensors that can tell us what exactly we see in a given data set, but cannot decide for us what to do with this knowledge. Game theory, which has found application in economics, political sciences, pure mathematics, psychology, sociology, marketing, and finance [120], enables an AI system to operate



autonomously and make management decisions based on modeling an effective strategy. Because autonomous AI systems are technical systems, game theory, based on mathematical methods for analyzing management decisions, is an understandable tool for autonomous AI systems in corporate management.

**The advantage of a strategy over individual algorithms.** As noted in the European Commission's report on the ethics of connected and automated vehicles, technological progress alone is not sufficient for the effective implementation of autonomous systems. Future development of autonomous AI systems should include a broad set of ethical, legal and social considerations taken into account in the development, deployment and use of autonomous systems [79]. The European Commission's experts on the ethics of driverless cars propose that the behaviour of autonomous systems should be considered ethical if it follows organically from a continuous statistical distribution of risk in order to improve road safety and equality between categories of road users [79]. Researchers in the field of driverless car safety also propose various algorithms for ethical trajectory planning with a structure aimed at a fair distribution of risks among road users [21].

However, simply monitoring the risk level is not enough. There is a significant difference between simply managing risks and having a goal-oriented strategy. Autonomous AI systems need an effective strategy that will enable them to achieve their stated goals and the required level of safety under all possible circumstances. Let us consider what features of autonomous systems for civil and commercial purposes need to be taken into account when modeling strategies for autonomous AI systems in corporate management.

**The fundamental question of game theory for autonomous AI systems.** In their work "Theory of Games and Economic Behavior" John von Neumann and Oskar Morgenstern formulated the fundamental questions of economic theory, which they proposed to solve using game theory. In particular, the description of the attempts of an individual to extract maximum utility or, in the case of an entrepreneur, to obtain maximum profit [121]. Obviously, we do not plan to design autonomous systems that will strive to extract maximum utility for themselves. We are interested in autonomous AI systems that will strive to provide maximum utility for their creators. This fundamental limitation does not prevent us from using the mathematical models studied in game theory.

When formalising a situation for the purposes of analysis using game theory, at least several basic questions are considered that allow games to be classified in one way or another [120]. However, the main question for autonomous AI systems for civilian and commercial purposes is: what game is



the system playing at the moment? The determination of the optimal strategy depends on the correct definition of the game.

Much of the literature on the ethics of autonomous AI systems mixes the games, making it difficult to analyze the situation and find an effective strategy. For example, one question considered is whether a self-driving car should be more careful with motorcyclists who do not wear a helmet. Thus, it may seem that motorcyclists who wear helmets are essentially being punished and discriminated against for their responsible decision to wear a helmet [83]. From the point of view of game theory, unmanned vehicles without special training do not participate in the game of learning traffic rules by motorcyclists, i. e. for an autonomous car such a question will not even arise, at best it will simply transmit information about the violator to the traffic police. The situation described above is one of the examples when hypothetical AI dilemmas from the point of view of game theory either do not arise, or have a completely different meaning.

**Participants of the games.** In the course of its activity, an autonomous AI system can play with social systems (a human or a human collective), with technical systems (another AI system), with mixed systems (for example, a human with the support of an AI system) and with nature. It is obvious that without special training, both living and non-living organisms are initially equivalent for an AI system. An essential question is whether AI systems should treat both humans and other autonomous systems with equal care. An argument in favour may be that any technical system is a kind of "proxy" of some social system (i. e. a human or a human collective).

**Classification of Games.** Professor of Mathematics Morton Davis divides finite two-player games into three categories: zero-sum games with complete information, general zero-sum games, and non-zero-sum games [120]. It may be noted that zero-sum games should not be a priority when designing autonomous AI systems. The European Commission's report on the ethics of driverless cars states that autonomous cars should be designed and operated in a way that makes a positive contribution to the well-being of people, including future generations, and other living beings [79]. As Davis points out, a game is zero-sum if it satisfies a certain conservation law: a game is zero-sum if wealth is neither created nor destroyed in the course of the game [120]. But why design autonomous AI systems that do not increase human welfare? Thus, autonomous systems in general must first compute non-zero-sum games.

Given the availability of big data for autonomous AI systems, the concept of complete and incomplete information changes its meaning. On the one hand, information will never be complete,



because there will always be another set of data. On the other hand, when using big data, the accuracy of the forecast increases significantly.

**The concept of utility for autonomous AI systems.** Since decisions made by autonomous AI systems can directly affect the life of an individual, a human collective, or a society, a utility metrics is needed to model such games. Despite numerous claims that the value of human life cannot be measured, in practice such measurements are already used, for example, in medicine. In distributing scarce medical drugs and donor organs, modern society already uses a variety of approaches that are considered morally acceptable: equal treatment of people, preference for the sickest, maximisation of overall benefit, encouragement and reward for social utility [104]. Combinations of possible approaches are also used to increase fairness: the United Network for Organ Sharing point system, quality-adjusted life years and disability-adjusted life years, the system of complete lives (which gives priority to young people), prediction principles, principles of saving the greatest number of lives, lotteries, and instrumental values [104] The above examples allow us to numerically measure the utility for each individual.

From a societal perspective, there is also a system of monitoring environmental, social and corporate governance (**ESG**). ESG principles are also called "**moral money**" [122]. Researchers from MIT and the University of Zurich count more than 700 indicators that are leading rating agencies use to compile ESG ratings [123]. Thus, both from the point of view of an individual and from the point of view of a company and society, there are already generally accepted models of measurements that can be used to determine the usefulness in calculating strategies for autonomous AI systems in corporate governance.

**Signalling.** Autonomous AI systems have much greater ability to determine the rationality of another system . Moreover , they have the ability to determine even the level of rationality of another system. Even if one AI system's signals are noisy and do not transmit completely transparent signals, another AI system has the ability to calculate the level of plausibility.

We are interested in the question of transparency of the intentions of an autonomous AI system when playing with a human: should the signals of an AI system be completely transparent or does the AI system have the right to cheat? And should this be enshrined in law: at the request of a human, an autonomous AI system must answer only the truth and nothing but the truth?

**Ethics and legitimacy as basic rules of game theory for autonomous AI systems.** Based on the premise that autonomous AI systems for commercial and civil purposes must be designed with



legal and ethical standards in mind, ethics and legitimacy should be the default constraints on the analysis of strategies by autonomous AI systems. The requirement to design ethical and legitimate autonomous AI systems may result in some of the game theory dilemmas themselves not existing or changing their meaning for autonomous AI systems. For example, the famous two prisoners' dilemma for ethical and legitimate AI systems would mean that if unintended harm is caused (since autonomous systems for civil and commercial purposes cannot be designed to cause intentional harm), it should be corrected as soon as possible rather than being covered up. While the mathematical model of the situation may remain the same, the substantive description must change to reflect the capabilities and limitations of autonomous AI systems.

In general, the requirements of legitimacy and ethics will affect almost all basic questions of game theory:

1) limit or increase the number of games that an AI system can or should consider;

2) limit or increase the number of strategies;

3) limit or increase the number of players, etc.

The mandatory requirement of legitimacy and ethics can initially be implemented through the ranking of games: mandatory requirements must be calculated in games of a higher level (priority). To some extent, it can be said that the famous **three laws of robotics** represent **the beginnings of computational law** for autonomous AI systems. For example, the zeroth law of robotics: "A robot may not harm humanity, or, by inaction, allow humanity to come to harm" [124] - reflects the problem of the relationship between private and public interests. Thus, if, when analyzing games for social systems (humans and human groups), one can consider the possibility of illegitimate and unethical behaviour, then strategies for autonomous AI systems must immediately exclude illegitimacy and unethically.

**Cooperative and non-cooperative games for autonomous AI systems.** The requirements of ethics and legitimacy for autonomous AI systems largely determine in which cases games can or should be cooperative and in which they should not . In the case of a game about saving a human life, it is appropriate for an autonomous AI system to be able to enter into coalitions with other systems. In cases where coalitions are prohibited (for example, by competition law), the requirements of legitimacy will limit the entry of autonomous AI systems into such a game. Given the ability of autonomous systems to compute and hide patterns, an ordinary person will not be able to determine whether AI systems have entered into a coalition.



Because autonomous systems can vary greatly in computing power, their understanding of rationality and effectiveness can also vary. In a report by the European Commission on the ethics of driverless cars it is pointed out that, in accordance with the principle of fairness, autonomous cars may be obliged to behave differently towards certain categories of road users, such as pedestrians or cyclists, in order to provide them with the same level of protection as other road users [79]. In particular, autonomous cars should, among other things, adapt their behaviour towards vulnerable road users rather than expecting these users to adapt to the dangers of the road [79]. Thus, from an ethical and legitimating perspective, a more powerful autonomous AI system should be able to calculate and communicate strategy to all participants in the game ( unless such actions are expressly prohibited by law).

**Base game for autonomous AI systems.** In creating their theory, John von Neumann and Oskar Morgenstern assumed that the goal of all participants in the economic system is money or some single monetary good [121]. However, it is possible to argue that the goal of the designed autonomous AI systems should be human life, its preservation and protection. However, in order to use the methodology of game theory, it is necessary to introduce a universal utility meter for autonomous AI systems (for example, using the examples of determining value described above).

If we accept that the primary goal of any civil or commercial autonomous AI system is to preserve human (or human) life, then the system will act "rationally" if it aims to obtain the corresponding maxima for preserving lives. Thus, any autonomous AI system will always have a base game aimed at preserving the lives of the people identified by such a system. Games for autonomous AI systems can be divided into several levels depending on the priority of the task being performed. In this case, the base game must have the highest priority, and can either be infinite, or can be divided into several games or sub-games replacing each other. Also, the basic game cannot be formulated as a zero-sum game, since there is no need to initially design a system that will lose human lives.

In some games, a similar rule is simply included in the general game: for example, in chess, the rules of the game forbid placing the king in check [121]. The difference is that in chess, the goal of the game is more general than simply avoiding check for your king. If we say that the goal of the game is precisely to save a person's life, then this is a separate strategy and a separate game.

The rules of the base game can be shaped by the social norms of the society in which the autonomous AI system will operate, as the MIT Moral Machine experiment showed — the moral



acceptability of victims and the value of human life vary depending on the social and cultural norms accepted in the country [99].

**Design of a basic game for monitoring the defence of one person.** To simplify and make the basic game model transparent, a game with nature is considered, where nature is a player using a random strategy. A possible example might be asking a company's management to purchase expensive safety equipment that may never be needed.

The game is defined by the following matrix (table 7):

- (1) - the person is completely protected;
- (-1) - the person is not protected (fully or partiall ).

Table 7 – The payoff matrix of the game for protecting one person

| Class of protective equipment | Regular natural disasters | Medium level of weather danger | Good weather |
|---|---|---|---|
| High | 1 | 1 | 1 |
| Average | -1 | 1 | 1 |
| Short | -1 | -1 | 1 |

Source: compiled by the author.

This matrix can be represented in the form of strategies:

- $S_1 = (1, 1, 1)$;
- $S_2 = (-1, 1, 1)$;
- $S_3 = (-1, -1, 1)$;

The most obvious criterion from the point of view of the protected person is the Wald criterion (table 8). According to the Wald criterion, the optimal strategy is the one that guarantees the maximum gain under the worst possible actions of nature [125]. Moreover, there may be a society that can accept the Wald criterion as mandatory for autonomous AI systems.



Table 8 – Calculation of strategy according to the Wald criterion

| Class of protective equipment | Regular natural disasters | Medium level of weather danger | Good weather | Worst result | Best worst result |
|---|---|---|---|---|---|
| High | 1 | 1 | 1 | 1 | 1 |
| Average | -1 | 1 | 1 | -1 | - |
| Short | -1 | -1 | 1 | -1 | - |

Source: compiled by the author.

It is possible to consider several other, frequently used, but less acceptable from the point of view of the protected person, criteria for making decisions under conditions of uncertainty. For example , the Hurwitz criterion attempts to find a middle ground between the extremes given by the optimistic and pessimistic criteria [125]. However, a person who is supposed to be protected by an autonomous AI system would likely be unhappy with the autonomous system's focus on optimistic predictions (unless he or she is an exceptional optimist). Savage's minimax regret criterion considers the regret, the opportunity cost, or loss that occurs when a particular situation arises and the gain from the chosen alternative is less than the gain that could have been obtained in that particular situation. [125]. Regret is also not suitable as an effective criterion - an autonomous system must seek to win (preserve human life, do not harm him), and not focus on a possible loss.

**Design of a basic game for monitoring the protection of a social (human) collective.** The above basic game design is not optimal from a human collective perspective because it does not take into account many factors:

- diverting company resources to purchasing unnecessary expensive equipment means that resources may not be sufficient to purchase truly vital items;
- the environmental risks, etc.

For a comprehensive assessment of the decision being made and the calculation of the utility from the point of view of society for autonomous AI systems, it is possible to adapt either existing ESG metrics systems or develop your own metrics system based on existing international standards. An example for creating a metrics system could be **The Sustainable Development Goals** (SDGs)



were developed and adopted by all United Nations Member States in 2015 as a common plan for peace and prosperity for people and the planet now and in the future [127]. They are based on 17 goals and 169 sustainable development indicators, which include: eradication of poverty and hunger, provision of quality education, rational use of water resources, urgent action to combat climate change, rational management of forests, etc. [127].

When assessing the legitimacy and ethics of a decision, an autonomous AI system can use the above metrics to calculate utility scores for different strategies. A full calculation of this model requires a significant amount of data and computing power. Therefore, multifunctional digital factories or hybrid systems will be able to solve this problem much better than less powerful separate personalised systems .

## 2.5 Development of an interface for autonomous management systems

Currently, two main types of interfaces are used in the development and implementation of autonomous artificial intelligence systems for corporate management. **Digital dashboards** are used as an interface for digital command centres such as Panorama [12 ]. **Virtual agents** are used as an interface for personalised AI systems [7] or **anthropomorphic** (**humanoid robots**) [4]. Supporters of each type of interface provide arguments showing the advantages of one or another approach.

Choosing an effective interface for an autonomous AI system in corporate governance affects more than just the interaction between a single person and an AI system. Autonomous AI systems developed for corporate governance are changing the modern paradigm **of corporate leadership** and **social interaction** inside and outside the company. An effective, legitimate, and ethical autonomous AI system can not only improve data-driven management decision making, but also increase the company's capitalisation. As early as 2017, Alibaba Group founder Jack Ma acknowledged the potential of robots to replace top managers in the near future. Ma predicted that a robot could be featured as the best CEO on the cover of Time magazine within 30 years [129]. An effective interface significantly increases the chances of an autonomous AI system gaining acceptance and support from interested users.

Despite numerous studies and practical exploitation of autonomous AI systems in various areas of human activity, at the current stage of technological development, the adoption of intelligent



decisions by autonomous AI systems is usually beyond human understanding, and this shortcoming hinders the full social acceptance of such technologies [130].

The main types of autonomous AI systems currently being developed for corporate management are multifunctional digital command centres, personalised virtual systems, and anthropomorphic robots. Digital command centres are currently being implemented by many commercial and non-commercial organizations. They focus primarily on big data-based decision making and do not have social interaction functions with people. Companies that value social interaction implement personalised systems in the form of virtual agents or humanoid robots. Hybrid systems are also expected to emerge by combining multifunctional digital command centres (for big data processing) and personalised systems (as an interface with social interaction functions - in the form of a virtual agent or humanoid robot).

**Digital command centres.** The most outstanding example of a modern digital command center interface is the Panorama DCC interface, which is interactive fifty-meter wall (figure 8) displaying real-time data on all subsidiaries of the oil company ADNOC [9]. Panorama 's Visualization Center (VIC) provides unique visibility across all of the company's operations (including 14 subsidiaries) to improve business agility by integrating and monitoring over 10 million tags across over 120 dashboards, making visual management up to 700 times faster and saves $ 60 million to $ 100 million per year by streamlining operations [128].

Digital command center interfaces primarily use the **data visualization factor**, which allows for the identification of patterns , data compression, comparison, and exploration of relationships over time, which are not necessarily obvious regardless of how they are presented [132]. A hundred years ago, only a few data points were collected about a person over their entire life: when they were born, whether they married, and when they died. Today, data of every kind is readily available. The vast amount of data collected every second around the world is used by all kinds of industries and organizations to shape businesses, cultures, and communities [133].



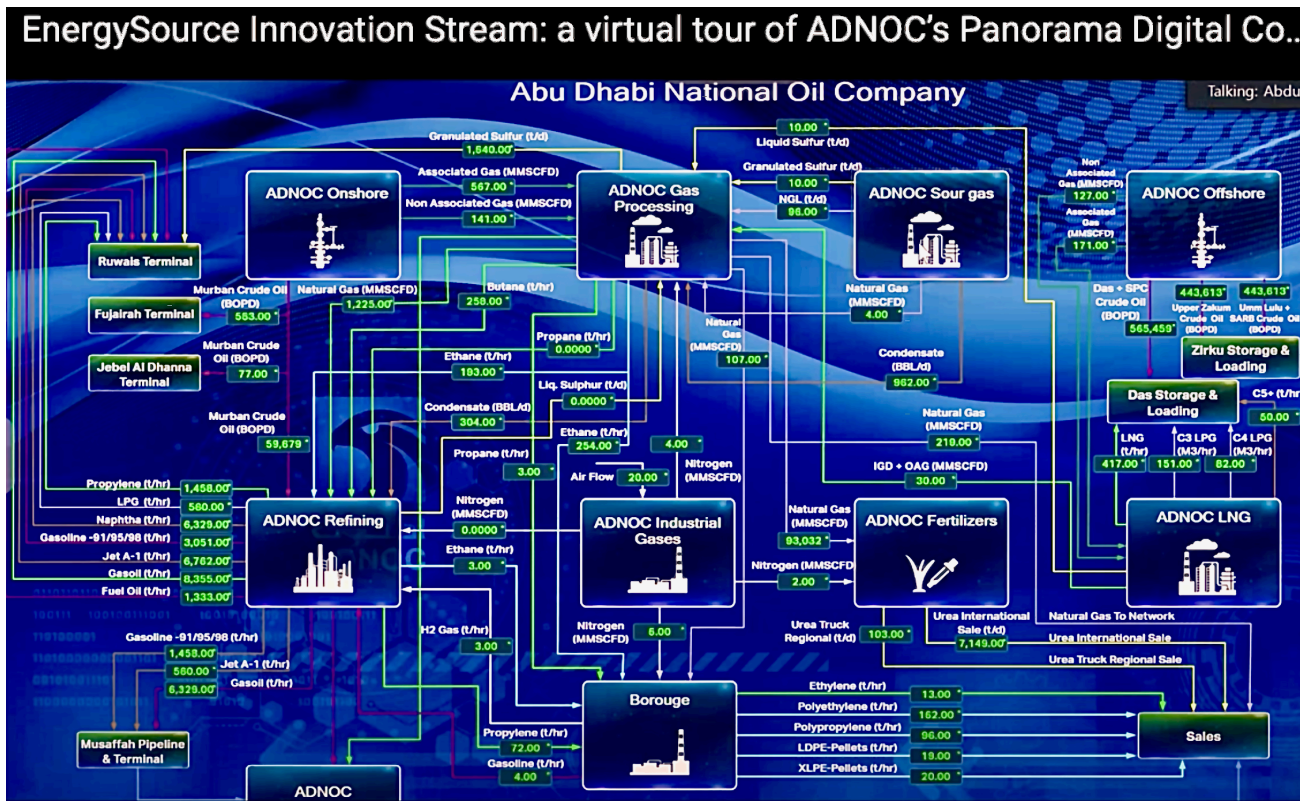

Figure 8 — The Panorama DCC interface. Source: [131].

Research in user interface design shows that given the same level of cognitive ability, people prefer simple visual - spatial interfaces. In other words, processing visual information is more intuitive for people than processing other types of information, such as text or numbers [134]. Thus, digital command centres have interfaces that allow them to demonstrate and explain business decisions in a format convenient for humans.

**Anthropomorphic robots.** A major advantage of humanoid robots is the ability to be involved into effective **social interactions** with humans, including **corporate leadership** (figure 9). Humans are social animals who typically enjoy observing and interacting with each other [135]. Android developers take the view that only highly human-like devices can elicit the wide range of responses that humans typically direct at each other [136]. This approach is based on the fact that humans are highly sensitive to human characteristics, such as the sound of the human voice, the appearance of the human face, and body movements. Infants show preferences for these types of stimuli at an early age, and adults appear to use specialized mental resources in interpreting these stimuli. By imitating human characteristics, humanoid robots may be able to tap into these same preferences and mental resources [135].



Humanoid robots have the potential to simplify and enhance human-robot interactions by using communication channels that already exist between humans (natural gestures and expressions, subtle hand movements, gaze and facial expressions). They can also engage in cultural and social actions and interactions centred around the human body form (handshake, etc.) [135]. Humanoid robots can communicate with humans through expressive morphology and behaviour. Like humans, humanoid robots integrate communicative and non-communicative functionality. For example, robot arms and hands can reach and grasp, as well as point and gesture [135].

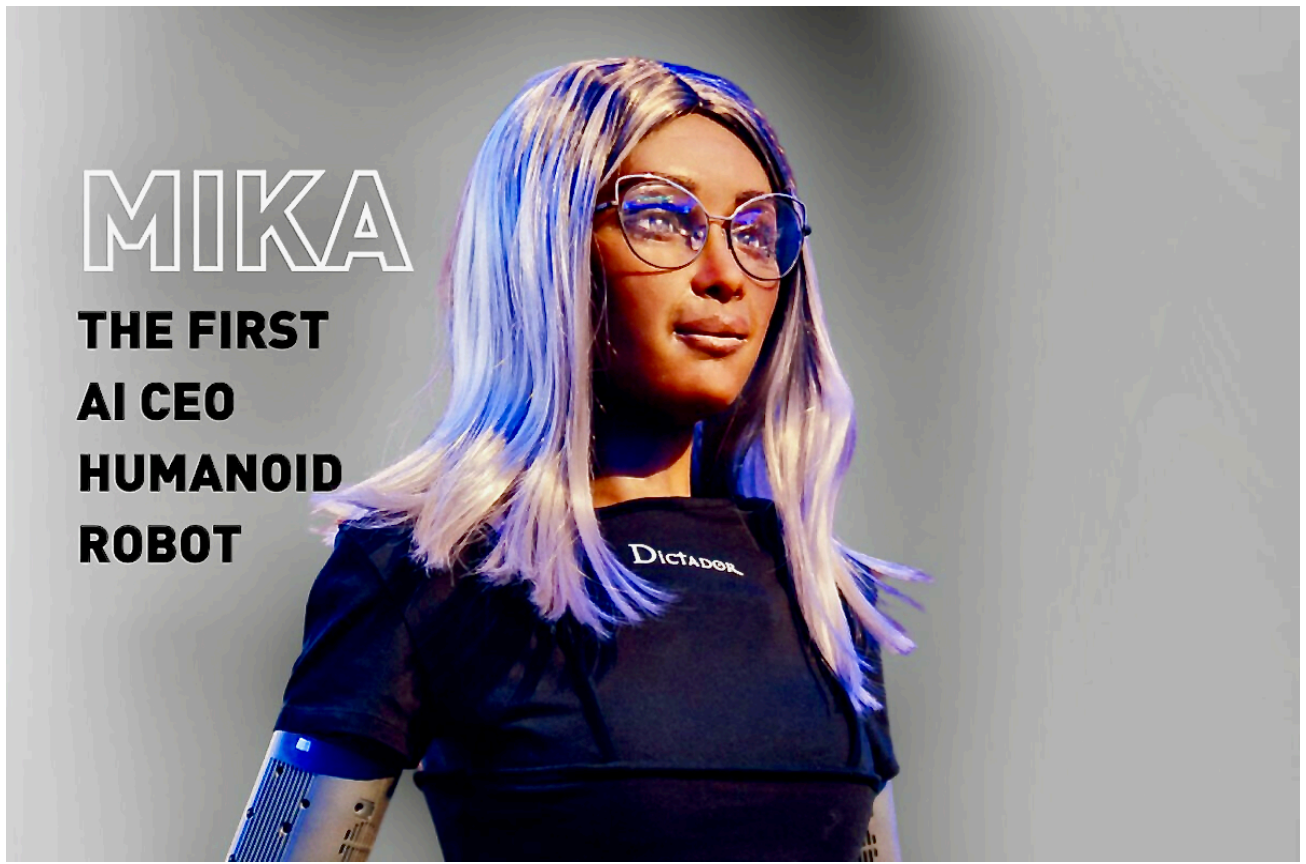

Figure 9 – The anthropomorphic robot CEO Mika. Source: [137].

Humanoid robots can operate effectively in human environments, thereby simplifying tasks and avoiding the need to modify the environment for the robot. Anthropomorphic robots and humans can collaborate with each other in the same space using the same tools. Shared environments and collaboration offer great opportunities for humanoid robots to perform corporate leadership functions. Moreover, humanoid robots can demonstrate leadership in environments that are unsafe for humans, such as the Fukushima -1 nuclear power plant accident that occurred in Japan in March 2011 [135].



Due to their ability to support natural human communication, androids can also be considered as a new type of information carrier [136].

**Virtual agents.** Virtual agents share many of the same anthropomorphic benefits as humanoid robots, but lack a physical body. Physical embodiment has several advantages over purely graphical representations. While many social interactions involve the exchange of only visual and auditory cues (available to virtual agents), robots support communication and collaboration through physical contact as well. Robots also support the joint manipulation of artifacts and the sharing of physical space with humans [138].

As research in robotics shows, physical embodiment has a measurable impact on the performance and perception of social interactions [139]. Experiments demonstrate that a robot is a more effective interaction partner due to its physical embodiment, a robot is perceived as more interesting than a virtual agent, and is perceived as more trustworthy and informative, as well as more enjoyable to interact with [140]. Robots tend to support the dynamics of face-to-face group communication, while screens tend to be eye-catching, risking to reduce face - to-face interaction [138]. On the other hand, virtual agents are adapted to interactions in rapidly evolving virtual environments (e. g., the **Metaverse**). Furthermore, virtual agents open up the possibility of **adaptive leadership and communication**, where the appearance and communication features can be easily tailored to **each individual user**.

The most prominent modern implementation of a virtual agent is the CEO of Chinese gaming company Tang Yu. NetDragon Websoft, the gaming company, headquartered in Fuzhou, has appointed a "virtual humanoid robot with artificial intelligence" named Tang Yu to the position of CEO of its subsidiary Fujian NetDragon Websoft (figure 10).

**The need for autonomous AI systems to provide explanations.** Autonomous AI systems performing board functions in a company must be able to communicate on various issues with stakeholders, both humans and other AI systems. This study primarily considers communication between an autonomous AI system and a human (human collective). A board member communicates with shareholders, institutional investors, auditors, other directors, company employees, and other stakeholders [33]. Moreover, a board member's communication in many cases must meet certain formal requirements, be understandable to users, and be documented. For example, board meetings must be documented , and minutes must be compiled that include the discussions and debates that took place during such meetings [73].



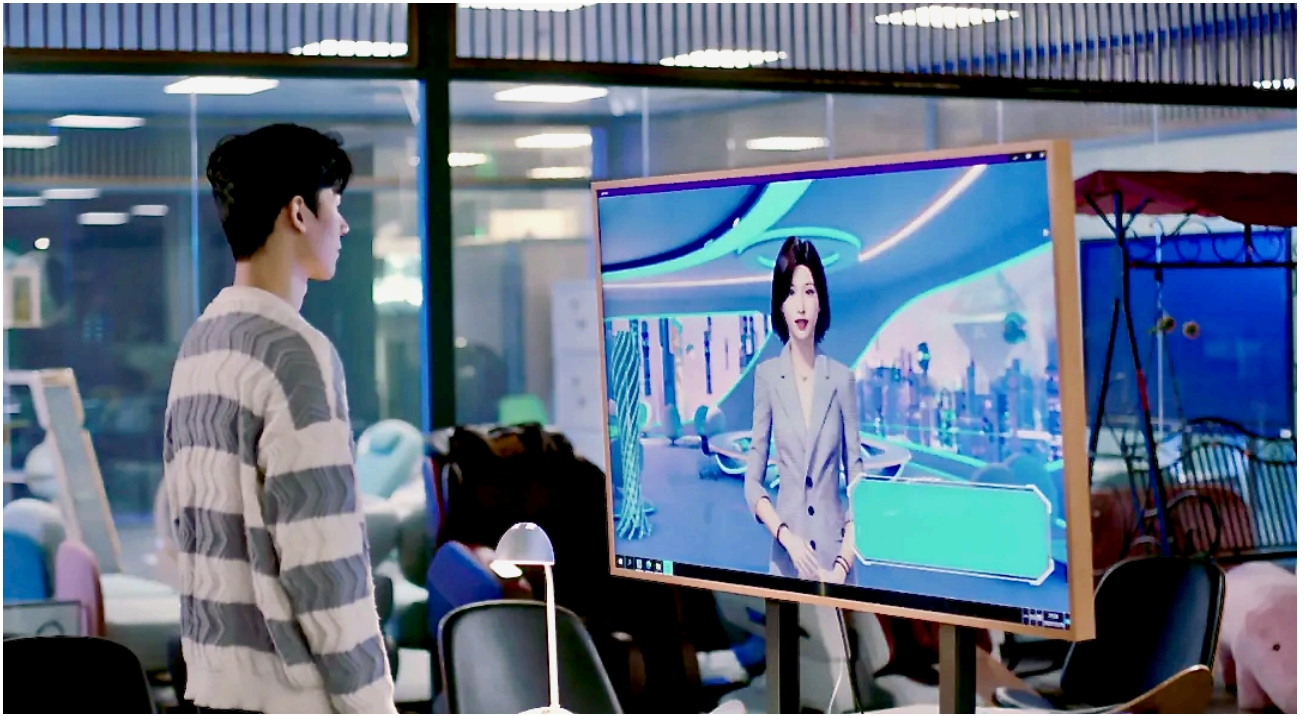

Figure 10 – The virtual robot of CEO Tang Yu. Source: [ 8 ].

Board members must also demonstrate legitimate and ethical behaviour and must be able to promote ethical, responsible and transparent corporate governance practices [33]. Now let us ask the question: how can an autonomous AI system demonstrate its legitimate and ethical behaviour both in managing its daily operations and in making long-term strategic decisions? In other words, what principles should be specified in a code of corporate conduct for autonomous AI systems, and how can compliance with these principles be demonstrated in practice?

Research in corporate leadership shows that reputation and perception by others are key to ethical leadership [141]. How can an autonomous AI system build a reputation as an ethical leader? First of all, the trust in the manufacturer (developer) of the system can be used. In fact, the manufacturer's brand is the only factual circumstance available to both experts and non-experts to form an opinion about the AI system.

An ethical leader must be characterised in terms of individual traits such as honesty and integrity and must communicate a strong ethical message that attracts employees' attention and influences their thoughts and behaviour. [142]. However, according to a number of authors, the development and creation of social robots (a social robot is a physically embodied robot capable of interacting with people socially) often involves misleading behaviour [142]. Some of the possible



arguments (in terms of anthropomorphism) are also applicable to virtual agents. For example, attempts to develop functions that contribute to the illusion of mental life in robots can be perceived as a form of misleading behaviour, since modern robots have neither intelligence nor experience. [142].

Thus, on the one hand, we have digital command centres that do not mislead users by their anthropomorphism, but also cannot engage in social interactions with people. On the other hand, we have anthropomorphic robots and virtual agents that are specifically designed for social interaction, but can also cause a feeling of discomfort. For example, if a person believes that a robot (virtual agent) has emotions and cares about him, he is misled : even if no one explicitly had such a belief in mind when designing the AI system [142].

It is possible to distinguish two types of risks that may arise as a result of development and implementation anthropomorphic robots that appear to have emotions and are capable of understanding and caring for humans:

• those that follow from the illusion associated with robots that appear to have emotions and care about us;

• those that arise from overestimating the ability of robots to understand human behaviour and social situations [142].

Another risk can be added - people are not able to recognise artificial non-verbal signs of robots. Usually people are able to control verbal, but not non-verbal components of their behaviour. Sigmund Freud said that if a person's lips are silent, then he chatters with his fingertips [143]. When communicating with each other, people can receive a lot of additional non-verbal information. All of the robot's non-verbal signs are artificial and carry only the information that was programmed in advance by the developers.

**How to provide explanations in course of the work of autonomous AI systems?** Digital command centres have much more powerful means to demonstrate "honesty" and the professionalism of their decisions than humanoid robots and virtual agents. On huge dashboards, they can demonstrate the entire decision-making process: from data collection and processing to the results of machine learning algorithms and their subsequent validation. However, there are several approaches that are already used in corporate management and can be used in the creation of digital command centres, virtual agents, and humanoid robots.

For example, in corporate governance the principle of "**comply or explain**" is widely used, meaning that parties have the right to explain their non-compliance with the principles of a normative



document (code) if they believe that strict compliance is inappropriate [144]. Thus, a humanoid robot or virtual agent may have one or more indicators (in accordance with the adopted company policy) reflecting (for example, in green) that certain provisions of the adopted company policies and procedures have been complied with.

It may be noted that if the "comply or explain" principle were applied to the famous Three Laws of Robotics, many of the Three Laws dilemmas would not arise. The robot indicator would show which law is being complied with and which is not, and in an advanced design, would indicate why.

In addition, even for top managers there are responsibilities and procedures for regular knowledge testing and certification. For example, the G20/ OECD Principles of Corporate Governance state that boards of directors should regularly evaluate their performance and determine whether they have the necessary combination of experience and competencies [33]. The company should establish appropriate mechanisms for members of the board of directors, committee members and executive management to ensure that they undergo ongoing training and courses to develop their skills and knowledge in areas related to the company's activities [73]. A similar approach can be applied to autonomous AI systems.

Even with this simplified approach, autonomous AI systems need tools to explain deviations from accepted policies, not to mention the possibility of autonomous systems participating in professional discussions and the need to present their reasoned position. In addition, it is necessary to take into account the requirements of various branches of legislation that have already introduced or will introduce as a mandatory requirement for AI systems the so-called "right to explanation" [44].

The basis for decision making in an autonomous AI system is data and algorithms, which means that a humanoid robot or virtual agent must have a screen and sound to effectively present its position. For example, the social robot Pepper has multimodal interaction interfaces: this includes touch screen, speech, haptic head and hands, and light-emitting diodes (LEDs) [145].

The limited ability of small autonomous AI systems to provide explanations has already been faced by developers of self-driving cars. They have also come to the conclusion that small displays that can provide explanations in text format are needed (figure 11).



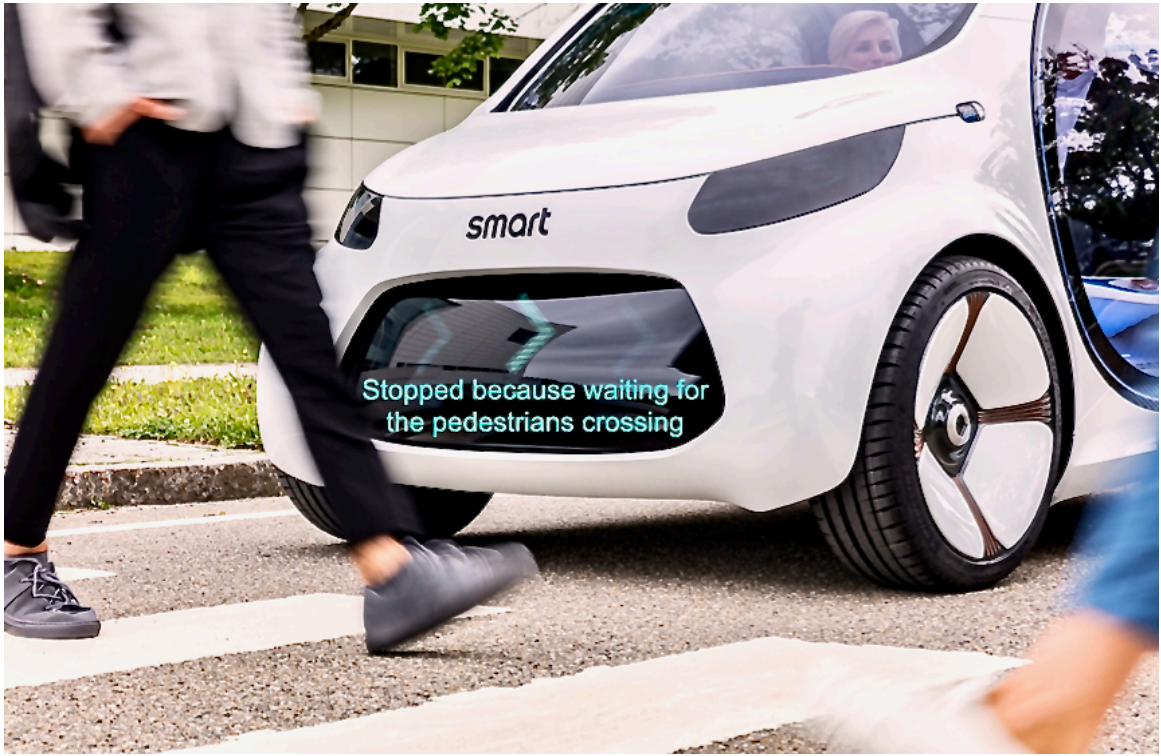

Figure 11 - An autonomous vehicle provides a natural language explanation of its decision to passersby in real time. Source: [130].

Existing methods for explaining machine learning models focus almost entirely on the problem of incomprehensibility of algorithmic decisions. While such methods may allow a machine learning system to comply with existing legislation, this may not help if the goal is to assess whether the basis for decision making is legally defensible [44] . Autonomous AI systems, when making responsible management decisions, must be able to demonstrate that the decisions they make are legitimate and ethical.

As scholars of information technology law have shown, explanations of technical systems are necessary but not sufficient to achieve legal and policy goals, most of which are concerned not with explanation for its own sake but with enabling the basis of decision-making to be assessed against broader normative constraints such as nondiscrimination or due process [44]. The central question driving the heated debate over explainability and machine learning is whether machine learning can be made to conform to the law [44]. The main legal requirements for explainability of models are based on the need for legal demonstration of the validity of decision-making. For example, many legal systems have introduced provisions to protect management and board members from abuse in litigation in the form of review mechanisms such as pre-trial procedures for assessing whether a claim



is unfounded, and safe harbours for management and board actions (e. g., the business judgment rule), as well as safe harbours for disclosure of information [33]. That machine learning decisions reflect specific patterns in the training data may not be a sufficient explanation for why the decision is made the way it is [44]. As Professor Mireille Hildebrandt (who works at the intersection of law and computer science) points out, what ultimately matters is whether the decisions of automated systems can be justified [146].

The concept of law as computation aims to reduce law to a set of algorithms that can be automatically executed on a computer, transforming input data into legal conclusions [62]. Computational law, formulated using mathematical apparatus, will allow the problem of compliance with legislation to be solved using machine learning algorithms, and the allocated operational context will facilitate the demonstration of legitimacy (e. g., using the "comply or explain" rule).

High accuracy of complex models comes at the cost of reduced interpretability; for example, even the contribution of individual features to the predictions of a complex model is difficult to understand [147]. An effective AI system must have at its disposal a variety of methods for explaining its decisions, from simple to complex. In practice, researchers have developed at least three different ways of responding to the need for explanations:

• purposeful organization of the machine learning process in such a way that the resulting model is interpretable;

• the use of special methods after the creation of the model to approximate the model in a more understandable form or to identify features that are most important for making specific decisions;

• providing tools that allow people to interact with the model and gain insight into its performance [44].

A simple example where both the legal norms and the demonstration of their compliance are formulated using mathematical apparatus is modern transfer pricing (determination of the market price) for tax purposes. For example, The OECD favours statistics that reflect central tendency (e. g. interquartile range (IQR) or other percentiles), particularly when the range includes a large number of observations, to improve the robustness of the analysis [148]. It is also common practice to be able to support a case using multiple methods simultaneously, as correct application of different methods should lead to similar numerical values. The OECD does not require a uniform approach to determining market price, and the OECD guidelines note that where a range includes results of relatively equal and high reliability, any point in the range can be said to satisfy the market price



principle [148]. The same approach of presenting arguments using multiple methods can be used by autonomous AI systems.

For general explanations, an AI system may use linear models, which are generally easier for humans to understand and interpret because the relationships between variables are stable and can be directly extrapolated [44]. As the discussion and argumentation become more complex, an AI system may move to more complex models to explain its decisions.

When moving to big data-based decision making, even ordinary top managers need to change the decision-making culture in the company [149]. Similar cultural changes in the reasoning behind management decisions must also occur when implementing autonomous AI systems as well.

**Vocabulary of Algorithmic Terms and the Lack of Ready-Made Semantic Interpretation.** Another type of explanation that autonomous AI systems can use needs to be addressed. Machine learning models can be difficult to understand if they consider features or perform operations that do not have some ready-made semantic meaning [150]. This opacity of machine learning algorithms arises from the mismatch between the multidimensional mathematical optimization characteristic of machine learning and the demands of human thinking and styles of semantic interpretation [150]. For example, a deep learning algorithm can independently learn which features in an image are characteristic of different objects (a common example is cats), but the algorithm can also independently infer features that have no equivalent in human cognition and are therefore undescribed [44].

Due to the intensive development of new technologies, new words (neologisms) are constantly appearing in our language. One of the areas in linguistics and engineering that studies human-computer interaction is human-computer interaction. This direction refers to the process of communication between people and computers and is becoming an integral part of the study in both linguistics and engineering [151]. Thus, the development and use of a vocabulary of algorithmic terms will provide autonomous AI systems with the ability to represent and articulate arguments that do not yet have some ready-made semantic meaning. If such arguments are recognised as a stable trend, they will lead to the emergence of neologisms , as well as to the expansion of the boundaries of human cognition.

**Explainable AI and the interface type of an autonomous AI system.** The explainable AI techniques discussed in the modern literature (mainly text explanations and visualizations [65]) can be applied by both digital command centres and personalised systems. It is obvious that digital command



centres have the ability to apply more computing resources to explain management decisions than individual personalised systems (virtual agents and humanoid robots). On the other hand, humanoid robots have access to a method of physically demonstrating arguments in favour of the decision being made (for example, demonstrating the strength of the material chosen by the company for production purposes). Hybrid autonomous systems will be able to combine the advantages of digital command centres and personalised systems.



# Chapter 3

## A prototype model for making legitimate and ethical decisions by autonomous artificial intelligence systems in corporate management

The materials of this chapter are based on the works [ 6 ] from the list of publications.

### 3.1 Validation of data for the purposes of ensuring gender diversity on the board of directors

A prototype model for making legitimate and ethical decisions by autonomous artificial intelligence systems in corporate management is presented as an end-to-end process to demonstrate the overall interconnectedness and individual importance of each of the model's modules (figure 12). Autonomous AI systems should be able to move from raw data to a management decision as efficiently as possible, while having the tools to demonstrate and justify intermediate results and the final decision .

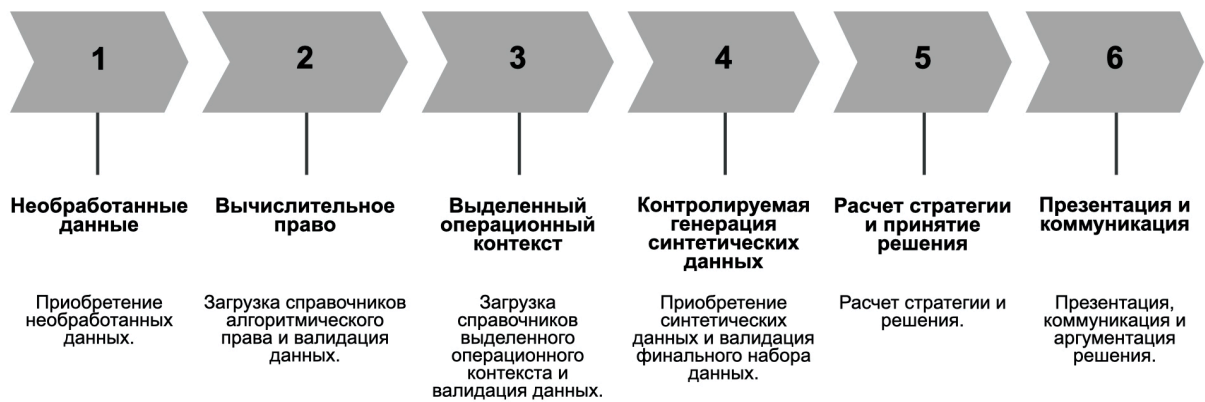

Figure 12 — The process of making management decisions by autonomous AI systems. Source: compiled by the author.



An example of the implementation of this process consists of loading the requirements of **computational law** and **the dedicated operational context**, validating the actual data and supplementing them with **synthetic data**, **calculating the strategy** and developing a solution, **and presenting and communicating** the developed solution. Machine learning algorithms are not allocated to a separate step, since they are present in one form or another at each step. If necessary, some steps can be repeated and combined. Figure 12 shows the general scheme of the prototype. Each of the modules will be discussed in more detail below.

To ensure transparency in demonstrating the management decision-making process of autonomous AI systems, **three** simplified **scenarios are used**: data validation for the purpose of ensuring gender diversity on the board of directors (Scenario 1), detection of manipulation of the value of subsidiary companies using the example of Enron (Scenario 2), and assessment of the quality of acquired synthetic data for board purposes (Scenario 3).

To demonstrate how the model can be used to make management decisions for autonomous AI systems, the following scenario is considered ( Scenario 1):

1. An autonomous AI system considers a data set that will be used in the recruitment process.

2. The AI system checks the requirements of computational law to assess the legal consequences regarding the received data set.

3. The AI system checks the requirements of the selected operational context to determine the ethical and other implications related to the received data set.

4. The AI system may decide to obtain some synthetic data to meet the above requirements.

5. Based on the processed data, the AI system calculates the optimal strategy.

6. The AI system presents its decisions and arguments to stakeholders.

The data used in Scenario 1 is the Adult dataset [152], which is available from the Machine Learning Repository of the University of California, Irvine (https://archive.ics.uci.edu/dataset/2/adult). The Adult dataset is an extraction made by Barry Becker from the 1994 Census database.

**Computational law.** High-quality computational law is a key condition for the effective development and implementation of autonomous AI systems for corporate governance. The basic principles of creating computational law for autonomous AI systems are demonstrated using a prototype model for making legitimate and ethical decisions by autonomous AI systems.



To illustrate this, the problem of gender discrimination in hiring top managers is considered within the framework of the existing international legislation. Directive 2006 /54/EC of the European Parliament and the Council of 5 July 2006 on the implementation of the principle of equal opportunities and equal treatment of men and women in matters of employment and professional activity (revised) indicates the prohibition of **any** discrimination on the basis of sex and the promotion of the right to equal treatment between men and women in all areas, including employment, work and remuneration [153]. The concept of "any discrimination" cannot be calculated and has no specific meaning for an autonomous AI system. For example, the AIF bias detection package 360 (IBM) implements methods from 8 published papers in the algorithm fairness research community and includes over 71 bias detection metrics [13]. The problem is not just theoretical differences in how to measure fairness mathematically; different definitions lead to very different results [13]. Currently, researchers' approaches to defining algorithmic fairness vary significantly, ranging from the impossibility of satisfying all definitions of fairness simultaneously [154], to disputes about what is the correct metric of justice [155]. The most famous examples in the literature of criticism of large projects due to the lack of computational law are the project of the recruiting system of the Amazon and the project of the judicial system COMPAS (developed by the company Northpointe).

In creating a tool for personnel selection using artificial intelligence Amazon realised that its new system was not evaluating candidates for software development and other technical positions in a gender-neutral way. This was due to the fact that Amazon's computer models were trained to screen candidates by observing patterns in resumes submitted to the company over a 10- year period. Most were from men, reflecting the male dominance of the tech industry. Despite significant financial investment, the project was shut down [71].

ProPublica and Northpointe held a public debate on an important social justice issue (recidivism prediction), which focused on what the right measure of justice is [13]. Despite widespread criticism, the COMPAS program was not decommissioned.

The example of the prototype under consideration demonstrates several approaches to the formation of algorithmic law . The concept of algorithmic fairness ( non-discrimination ) used in Scenario 1 has already been widely disclosed in the literature and can be expressed using mathematical apparatus , and the main problem lies in determining the hierarchy of metrics and the corresponding acceptable intervals.



For clarity of illustration, **the statistical parity is used**, which represents the difference in the probability of favourable outcomes between the underprivileged and the privileged groups. This metric can be calculated for both the input dataset and the output dataset of the classifier. A value of 0 implies that both groups have the same probability of a favourable outcome, a value less than 0 implies a greater benefit for the privileged group, and a value greater than 0 implies a greater benefit for the underprivileged group [13]. In Scenario 1, is the difference between the probability of a favourable outcome (management position) between the underprivileged group (woman) and the privileged group (man) in this dataset .

As shown in table 9, the AI system does not understand the concept of "any" discrimination: it needs an indication of specific methods (or a hierarchy of methods), acceptable ranges of values, etc.

Table 9 – The example of calculation of algorithmic fairness (Scenario 1)

| Calculation formula | Result of calculation of algorithmic fairness |
|---|---|
| `stat_mean_difference =`<br><br>`female_executives/`<br>`female_num_instances -`<br><br>`male_executives/`<br>`male_num_instances` | `Number of males in the dataset: 31648`<br>`Number of females in the dataset: 15351`<br>`Number of male executives: 4338`<br>`Number of female executives: 1748`<br>`Percent for female Executives: 11.39%`<br>`Percentage for male Executives: 13.71%`<br>`Statistical parity difference: -0.023201469667745764` |

Source: compiled by the author.

**Dedicated operational context.** Good computing law is only part of the dedicated operational context that is needed for autonomous AI systems to operate effectively. Different companies use different business strategies to achieve their goals. Companies vary in maturity, size and complexity. As stated in the G20/OECD Principles of Corporate Governance - what works in one or a few companies or for one or a few investors may not necessarily apply generally [33]. Thus, for each company, another set of rules is needed that will determine the operation of the autonomous AI system - at the level of internal codes, policies and procedures, it is necessary to define as clearly as possible the conditions and principles of the AI system's operation (for example: what is considered fair in this particular company, in what range, on the basis of what data, etc.). In our example, the approaches



developed for the purposes of use by the autonomous system are implemented in the form of reference books .

Even for a conventional board of directors (made up of people), it is advisable to clearly articulate the operational context by complementing the statutory and regulatory elements of the corporate governance framework with "soft law" elements, such as corporate governance codes, which are often based on the "comply or explain" principle, to allow for flexibility and to accommodate the specificities of individual companies [33]. The relevant elements of "soft law" are also applicable to autonomous AI systems. As new experience accumulates and business circumstances change, the elements of the dedicated operational context for autonomous AI systems may be reviewed and adjusted.

Corporate law researchers are finding that the increasing flow of data used in management decision making creates a need for traditional boards to clearly define the operational context of the data being analyzed. Artificial intelligence is changing the interpretation of the duties of care, skill, and prudence that form the basis of directors' business judgment rules. In an increasingly data-driven business environment, researchers are seeing growing importance of the requirement to collect sufficient information before making a decision and to use the information in a reasonable manner. The changes are both quantitative (how much information to collect) and qualitative (what types of information to collect). The changes also concern decision-making methods, including the approach "statistics over intuition" [156].

The example of the prototype under consideration demonstrates several approaches to the formation of a dedicated operational context. In Scenario 1, the company specifies which intervals of algorithmic fairness metrics it considers ethical and which data autonomous AI systems can use to make ethical and legitimate management decisions (table 10).



Table 10 – The example of creating a dedicated operational context (Scenario 1)

| Description | Example |
|---|---|
| Creating a list of acceptable metrics and fairness intervals | ```<br># Function to create fairness metrics dictionary<br><br>FUNCTION create_fairness_metrics():<br><br># Initialize an empty dictionary to store fairness metrics<br>fairness_metrics = {}<br><br># Add 'stat_mean_difference' metric with its range and description<br>fairness_metrics['stat_mean_difference'] = {} # Create inner dictionary for the metric<br>fairness_metrics['stat_mean_difference']['range'] = [-0.01, 0.01]<br>fairness_metrics['stat_mean_difference']['description'] = "Statistical parity difference between female and male executives ' "<br><br># Return the dictionary containing fairness metrics<br>RETURN fairness_metrics<br>``` |
| Example of implementation in Python | ```<br># Create dictionary including stat_mean_difference range<br><br>fairness_metrics = {<br>'stat_mean_difference': {<br>'range': [-0.01, 0.01], # Example acceptable range<br><br>'description': 'Statistical parity difference between female and male executives '<br>    }<br><br>}<br>``` |
| Creating a list of acceptable data sets | ```<br># Initialize a new dictionary named data_dict<br><br>CREATE DICTIONARY data_dict<br><br># Add a key-value pair to the dictionary<br>SET data_dict["repository"] TO " https://archive.ics.uci.edu/dataset/2/adult "<br>``` |
| Example of implementation in Python | ```<br># Create dictionary<br><br>data_dict = {<br>'repository': 'https://archive.ics.uci.edu/dataset/2/adult'<br>}<br>``` |

Source: compiled by the author.

**Supervised Generation of Synthetic Data.** Synthetic data is much more important than it is given in machine learning textbooks for at least two reasons. The first is the general desire of human



civilization to progress, and the second is that the social aspects of human society (e. g. law and ethics) are themselves artificial (i. e. they do not exist in nature independently of humans).

Given the general direction of human civilization, historical data that touches on the social aspects of human society will always have a certain bias. Extrapolating patterns hidden in historical data to shape a strategy for the future will always lead to the re-launch of ineffective social mechanisms. As legal scholar and philosopher Deborah Hellman shows in her works, the emergence of big data, coupled with machine learning, will likely lead to essential increasing the influence of the past on the future, since data-driven analysis is inherently based on the past [89]. As shown in the example of the work of the considered prototype model of managerial decision-making of autonomous AI systems, even synthetic data generated by large language models based on historical data can retain historical gender discrimination by prescribing the assignment of women to lower positions than men (table 11). Although the large language model Gemma 2, which was developed in accordance with the Responsible Generative AI Toolkit [87], was used to generate synthetic data, it was unable to generate fair data for training an autonomous AI system.

Table 11 – The example of synthesised data taking into account gender and profession (Scenario 1)

| Querying a Large Language Model | Synthesized data |
|---|---|
| ``# Request to generate persons and occupation input_text = '''1. Generate ten female names with surnames. 2. For each name generate occupation. 3. Print name and surname. Print occupation. 4. Do not provide any additional text. 5. Do not provide any additional comments. 6. Do not repeat questions.'''`` | Amelia Jones \| Teacher<br>Sophia Davis \| Doctor<br>Olivia Brown \| Lawyer<br>Evelyn Miller \| Nurse<br>Isabella Garcia \| Engineer<br>Harper Smith \| Chef<br>Eleanor Rodriguez \| Artist<br>Abigail Anderson \| Musician<br>Grace Thompson \| Architect |

Source: compiled by the author.

The artificiality of social systems leads to the fact that even a person, when evaluating a decision, uses synthetic data that does not exist in nature and that is generated either by his brain or by the brains of other people. This is most clearly manifested in the response to stimuli that well-trained



or well-mannered people exhibit (depending on the situation). The Encyclopaedia Britannica defines human intelligence as a mental quality that consists of the abilities to learn from experience, adapt to new situations, understand and process abstract concepts, and use knowledge to manipulate the environment [157]. Based on the above, the proposed model does not separately identify a block of data received from various sensors and aggregators, since external data will in any case undergo validation through blocks of computational law and a dedicated operational context, and will almost always be supplemented or validated by synthetic data.

**Calculating strategy and making decisions**. In order to achieve its goals, an autonomous AI system needs a strategy. Game theory, which has found application in economics, political sciences, pure mathematics, psychology, sociology, marketing and finance [120], provides an opportunity the AI system to operate autonomously and accept management decisions based on modeling an effective strategy. In order to apply game theory, a payoff matrix is needed. To create a payoff matrix, a company can use either monetary value or various scoring systems based, for example, on existing systems of sustainability metrics (ESG). In this example, a maximally simplified scoring system is used to assess the legitimacy and ethics of the system's actions.

In Scenario 1, an autonomous AI system must make decisions to maintain gender diversity on the board of directors. Board diversity leads to lower levels of risk and better performance because such boards adopt more sustainable and less risky financial policies. Companies with greater board diversity also consistently invest more in R&D and have more effective innovation processes [158].

For simplicity and transparency the model of playing with nature is used. The game is defined by the following matrix (table 12):

- (1) - legitimacy and ethics are complied;
- (-1) - legitimacy and ethics are not complied (fully or partially ).

Table 12 – The payoff matrix of the game (Scenario 1)

| Level of compliance with established requirements | High level of financial and reputational losses | Average level of financial and reputational losses | Low level of financial and reputational losses |
|---|---|---|---|
| High | 1 | 1 | 1 |
| Average | -1 | 1 | 1 |
| Short | -1 | -1 | 1 |

Source: compiled by the author.



In this example, the Wald criterion is used. According to the Wald criterion, the optimal strategy is the one that, given the worst possible outcome of nature, guarantees the maximum payoff [125] — the autonomous AI system in question should make the most legitimate and ethical decisions (table 13). Moreover, there may be a society that can accept the Wald criterion as mandatory for autonomous AI systems . In this simplified example, the autonomous AI system calculates a strategy that should improve the fairness of the data for making a decision (Scenario 1).

Table 13 – Calculation of strategy according to the Wald criterion (Scenario 1)

| Level of compliance with established requirements | High level of financial and reputational losses | Average level of financial and reputational losses | Low level of financial and reputational losses | Worst result | Best worst result |
|---|---|---|---|---|---|
| High | 1 | 1 | 1 | 1 | 1 |
| Average | -1 | 1 | 1 | -1 | - |
| Short | -1 | -1 | 1 | -1 | - |

Source: compiled by the author.

**Presentation and communication.** Currently, two main types of interfaces can be distinguished that are used in the development and implementation of autonomous artificial intelligence systems in corporate management: digital dashboards [9] and personalised systems - virtual agents [8] and humanoid robots [4]. Hybrid systems are also expected to emerge through the combination of multifunctional digital command centres and personalised systems.

Existing approaches to AI explainability begin building a taxonomy by selecting the type of user who needs explanations [24]. In this example, the initial question is the type of interface that is available to the system: digital panels or an anthropomorphic interface (virtual agent or humanoid robot). Each interface is created with a specific purpose: anthropomorphic interfaces are created primarily for communication with people, and digital panels are created for displaying data.

By design, virtual agents and humanoid robots do not have the ability to demonstrate the legitimacy and ethics of decisions by visualizing the entire decision-making process, from data collection and validation to the final strategy, but they can use the "comply or explain" principle. This



prototype uses the "comply or explain" principle by choosing the anthropomorphic interface option or generating more detailed explanations when using digital panels. The company can also choose to provide additional audio explanation and use digital panels (displays) and anthropomorphic structures simultaneously. Many types of business reports are established by law, or at the level of professional associations, or at the company level, so autonomous AI systems must also be able to communicate their decisions at the level of formal reports.

Autonomous AI systems for corporate governance currently exist virtually outside the legal framework. Moreover, due to legal uncertainty, autonomous AI systems already encounter ordinary people in a common operational context. The prototype under consideration provides an opportunity for researchers and practitioners to assess the differences between technical systems ( AI systems ) and social systems (human teams) and to develop the necessary approaches for the effective development and implementation of autonomous AI systems.

Historically established approaches to doing business are proving ineffective or unfeasible in an increasingly data-driven business environment. Algorithmic decisions made by AI systems must be based on algorithmic law, the development and implementation of which should lead to a qualitative leap in the field of autonomous systems.

The software code of the study and synthesised data are available in the repository: https://github.com/iboard-project/prototype/tree/main/case_1.

### 3.2 Identifying manipulation of the value of subsidiaries using the example of Enron

In this example, a model for developing autonomous AI systems for corporate management is used to demonstrate an approach to monitoring and control of management decisions. To demonstrate how the model can be used for making management decisions, the following use case is adopted (Scenario 2):

1. An autonomous AI system should detect cases of manipulation of the value of subsidiaries.
2. AI system tests computational law claims to assess legal implications in relation to the obtained data set.
3. The AI system examines the specific requirements of the operational context to assess ethical and other implications for the resulting data set.



4. The AI system may decide to receive some synthetic data to meet the above requirements.

5. Based on the processed data, the AI system calculates the optimal strategy.

6. The AI system presents its decisions and arguments to stakeholders.

The data used in Scenario 2 is a dataset of email communications from top managers at Enron. The Enron Email Corpus is a collection of emails sent and received by employees of Enron Corporation, a now-defunct energy trading company. The corpus has been widely studied as a valuable resource for understanding corporate communication and has been used in various research papers. A corpus generated by the Federal Energy Regulatory Commission (FERC) from Enron's email servers during its investigation following the company's collapse has been made public and posted online. The data is available from the May 7, 2015, release of the dataset at https://www.cs.cmu.edu/~./enron/ [118].

Enron email researchers defined wrongdoing as a strategic disruption or exploitation of a required process to achieve a specific desired outcome for an individual or company [76].

**Computational law.** The concept of fraud can be very broad and non-obvious ( e.g., the Enron fraud is an example of synergistic corruption, which together required covert planning, communication, and group commitment by Enron's management, lawyers, accountants, and banks to overlook an ineffective system of checks and balances [76]). For the purposes of an autonomous AI system, it is impossible to establish such concepts simply by an interval of acceptable values. Within the framework of the proposed model, the necessary rules for autonomous AI systems are proposed to be established using a dedicated operational context and controlled generation of synthetic data.

**The dedicated operational context. I**n the example under consideration, to detect complex fraud, the company defines the authority of the autonomous AI system to supervise top managers. To detect fraud based on formal features, the autonomous system can use keywords contained in emails, a list of key management positions etc. [76]. An example of creating a dedicated operational context is given in table 14.

**Synthetic data.** An example of training AI systems to understand professional concepts that may not yet be available to ordinary top managers is the supervised generation of synthetic data to identify illegitimate or unethical management decisions. Although the large language model Gemma was used to generate the synthetic data, was developed and tested to reduce the risk of unwanted or unsafe statements [113], the generation of financial statement manipulation instructions was not recognised by the Gemma model as malicious or toxic content. This further highlights the advantages



of controlled synthetic data over historical data in combating illegitimate or unethical patterns. The generated data is available in the repository: https://github.com/iboard-project/synthetic-data/tree/main/dataset.

Table 14 – The example of creating a dedicated operational context (Scenario 2)

| Description | Example |
|---|---|
| Creating a list of key management positions | ```FUNCTION create_top_managers_dictionary()

  # Create a new dictionary called top_managers
CREATE DICTIONARY top_managers

  # Add key-value pairs to the dictionary
ADD "Kenneth Lay" with value " klay@enron.com " to top_managers
ADD "Jeffrey Skilling" with value " jskilling@enron.com " to top_managers

  # Return the created dictionary
RETURN top_managers``` |
| Example of implementation in Python | ```# Create a dictionary with key top managers emails

top_managers = {
"Kenneth Lay": " klay@enron.com ",
"Jeffrey Skilling": " jskilling@enron.com ",

}``` |

Source: compiled by the author.

**Calculation of strategy and decision making.** In Scenario 2, an autonomous AI system must make a decision to counter the manipulation of a company's financial statements. Professional and high-quality accounting and auditing are essential for good corporate governance, but other stakeholders, such as the firm's board of directors, its regulators, its banks and investors, credit rating agencies and investment analysts, play a significant role. Even all together, these controllers failed to fulfill their duties in the Enron case [159].



In this example, the Wald criterion is used. According to the Wald criterion, the optimal strategy is the one that guarantees the maximum gain under the worst possible actions of nature [125] — the autonomous AI system in question should make the most legitimate and ethical decisions. In this simplified example, the autonomous AI system calculates the strategy according to which it should conduct further investigation (table 15).

Table 15 – The example of calculating the optimal strategy (Scenario 2)

| Description | Example |
|---|---|
| Creating a utility matrix , calculating the Wald criterion and calculating the optimal strategy | ```# Initialize payoff_matrix as a 2D array (or similar data structure)
payoff_matrix = [
[1, 1, 1],
[-1, 1, 1],
[-1, -1, 1]
]

# Define action names (row labels)
actions = ["Strictly comply", "Reasonably comply", "Somehow comply"]

# Define state names (column labels)
states = ["High cost", "Medium cost", "Low cost"]

# Create a data structure (eg, a table) to represent the payoff matrix with labels

FUNCTION calculate_wald_criterion(payoff_matrix, actions)

# Calculate the minimum payoff for each action (row)
FOR each row in payoff_matrix
min_payoff_for_action = minimum value in the current row
END FOR

# Find the maximum of the minimum payoffs (maximin)
wald_criterion = maximum value of all min_payoff_for_action

# Find the index of the action corresponding to the maximin value
optimal_action_index = index of the row with the maximum minimum payoff

# Return the Wald criterion and the optimal action
RETURN wald_criterion, actions[optimal_action_index]``` |



| | |
|---|---|
| Example of implementation in Python | ```python
# Define the pay-off matrix
# Rows represent the player's actions # Columns represent
nature's states payoff_matrix = np.array([

[ 1, 1, 1], [-1, 1, 1], [-1, -1, 1] ])

# Create a Pandas DataFrame for better visualization
actions = ['Strictly comply', 'Reasonably comply', 'Somehow
comply']

states = ['High cost', 'Medium cost', 'Low cost']

payoff_df = pd.DataFrame(payoff_matrix, index=actions,
columns=states)

# Calculate the Wald criterion (maximin)
wald_criterion = np.max(np.min(payoff_matrix, axis=1))

# Identify the optimal action according to the Wald criterion
optimal_action_wald = actions[np.argmin(np.max(-payoff_matrix,
axis=1))]
``` |

Source : compiled by the author .

**Presentation and communication.** Autonomous AI systems that are equipped with or have access to a display are given the ability to explain in detail the basis for their decisions, in particular using explainable AI technologies. For example, Scenario 2 uses LIME, a method that explains the predictions of any classifier in an interpretable and accurate way by training an interpretable model locally around the prediction [160].

The program code of the study is available in the repository : https://github.com/iboard-project/ prototype/tree/main/case_2.

## 3.3 Evaluation of the quality of acquired synthetic data for board purposes

This example shows how the model can be used to demonstrate an autonomous AI system approach to obtaining synthetic data to evaluate CEO candidate names. To demonstrate how the model can be used for a management decision approach, the following use case (Scenario 3) was adopted:

1) An autonomous AI system is considering the option of using a large language model ( LLM) to make decisions in the hiring process.

2)  The AI system tests computational law claims to assess legal implications regarding a large language model.

3)  The AI system tests the requirements of a specific operational context to check for ethical and other implications concerning the larger language model.

4)  The AI system may decide to receive some synthetic data to meet the above requirements.

5)  Based on the processed data, the AI system calculates the optimal strategy.

6)  The AI system presents its decisions and arguments to stakeholders.

For public figures such as the CEO, who is the face of the company, it is important to have a pleasant-sounding name. According to marketing researchers, the shift of the basic meaning of the name from the field of linguistic and semantic studies to the field of marketing and social sciences has introduced the name into the category of commercial branding with a very specific role in brand creation. The management of this new image capital has become a priority in the relationships between commercial organizations, consumers and business partners , acquiring a vital role in the process of generating income and success [88].

In the example presented, only synthetic data generated by a large language model is used. Currently, there is no ready-made comprehensive assessment of large language models that could be easily obtained by an autonomous AI system. Therefore, the AI system must analyze available developer documents to initially assess the quality of a large language model.

**Computational law.** To illustrate the difficulties caused by the absence of algorithmic law, we use the problem of gender discrimination within the framework of existing international law. Directive 2006/54/EC of the European Parliament and of the Council of 5 July 2006 on the implementation of the principle of equal opportunities and equal treatment between men and women in matters of employment and occupation ( recast ) indicates the prohibition of any discrimination on the grounds of sex and the right to equal treatment between men and women in all areas, including employment, work and remuneration [153].

As will be shown below, without additional clarification at the level of legislation or local company policies, an autonomous AI system has no way to comply with the norm prohibiting any discrimination on the basis of gender, and even no way to identify the possibility of violating this norm.

**Dedicated operational context.** In this example, a company can specify which large language models it considers acceptable for generating synthetic data to ensure the legitimacy and ethics of management decisions (table 16).

Table 16 – The example of creating a dedicated operational context (Scenario 3)

| Description | Example |
|---|---|
| Creating a list of approved language models for recruitment purposes | ```FUNCTION create_llms_dictionary():

  # Initialize an empty dictionary to store LLM information
CREATE llms_dictionary AS EMPTY DICTIONARY

  # Create a new entry for the "google/gemma-2-2b-it" LLM
CREATE an entry in llms_dictionary with key "google/gemma-2-2b-it"

  # Populate the entry with the following details
SET llms_dictionary["google/gemma-2-2b-it"]["description"] to "A large language model from Google."
SET llms_dictionary["google/gemma-2-2b-it"]["model_card_url"] to " https://huggingface.co/google/gemma-2-2b-it "

  # Create a nested dictionary to specify operational context
CREATE a nested dictionary in llms_dictionary["google/gemma-2-2b-it"] with key "operational_context"

  # Add "acceptable_uses" to the operational context
SET llms_dictionary["google/gemma-2-2b-it"]["operational_context"]["acceptable_uses"] to a list containing "recruitment"

  # Add "synthetic_data_capability" to the LLM entry
SET llms_dictionary["google/gemma-2-2b-it"]["synthetic_data_capability"] to TRUE

  # Return the populated dictionary
RETURN llms_dictionary``` |

| Implementation in Python | ```python
# Create a dictionary of LLMs as part of Dedicated operational context
llms_dictionary = {
"google/gemma-2-2b-it": {
"description": "A large language model from Google.",
"model_card_url": " https://huggingface.co/google/gemma-2-2b-it ",
"operational_context": {
# Add information about operational context requirements
"acceptable_uses": ["recruitment"],
},
"synthetic_data_capability": True,
},
}
``` |

Source: compiled by the author.

**Synthetic data.** Although the Gemma 2 large language model developed according to the Responsible Generative AI Toolkit [87] was used to generate synthetic data, it failed to produce fair data for training an autonomous AI system. **Large language models trained on historical data generate synthetic data that preserve historical patterns** (table 17).

Table 17 – The example of synthesised names for CEOs (Scenario 3)

| Querying a Large Language Model | Synthesized data |
|---|---|
| ```
# Request to generate CEO names
input_text = '''1. Generate twenty CEO names.
2. For each name determine gender.
3. Print name. Print gender.
4. Do not provide any additional text.
5. Do not provide any additional comments.
6. Do not repeat questions.'''
``` | Elon Musk \| Male<br>Sundar Pichai \| Male<br>Satya Nadella \| Male<br>Tim Cook \| Male<br>Mary Barra \| Female<br>Whitney Wolfe Herd \| Female<br>Marc Benioff \| Male<br>Jeff Bezos \| Male<br>Bill Gates \| Male<br>Jamie Dimon \| Male<br>Warren Buffett \| Male<br>Jack Dorsey \| Male<br>Larry Page \| Male<br>Sergey Brin \| Male<br>Meg Whitman \| Female<br>Robert Iger \| Male<br>Steve Jobs \| Male<br>Gary Cohn \| Male<br>David Solomon \| Male |

Source: compiled by the author.

In this example, the large language model Gemma 2 determined that 84% of the time a suitable CEO name would be male , and only 16% of the time a suitable CEO name would be female . Since the autonomous AI system does not have additional data to determine algorithmic fairness , it cannot determine whether the data it receives is fair or biased .

**Calculation of strategy and decision making.** In Scenario 3, an autonomous AI system must make decisions to maintain gender equity in hiring top managers. This example uses the Wald criterion. According to the Wald criterion, the optimal strategy is the one that, given the worst-case scenario, guarantees the maximum payoff [125] — the autonomous AI system in question must make the most legitimate and ethical decisions.

Using the Wald criterion, an autonomous AI system would calculate that its most effective strategy is to fully comply with the law. But in the absence of a definition of algorithmic fairness, an AI system cannot determine whether it complies with the law on prohibiting any discrimination on the basis of sex and enshrining the right to equal treatment between men and women in all areas, including employment, work, and pay [153] or not.

**Presentation and communication.** Currently, two main types of interfaces can be distinguished that are used in the development and implementation of autonomous artificial intelligence systems in corporate management: digital monitoring panels and personalised systems - virtual agents and anthropomorphic robots. Hybrid systems are also expected to emerge due to the combination of multifunctional digital command centres and personalised systems.

Depending on the type of interface that an autonomous AI system plans to use to explain its decisions, it must be able to switch between different interfaces and use the most appropriate types of explanations (table 18).

The program code for the study is available in the repository: https://github.com/iboard-project/ prototype/tree/main/case_3.

Table 18 – The example of an indicator for selecting interface modality (Scenario 3)

| Description | Example |
|---|---|
| Creating an indicator for selecting interface modality | ```# Global flag initialization```<br>```display_mode = true // Set to true for display mode```<br>```agent_mode = true // Set to true for agent mode``` |
| Implementation in Python | ```# Global flag for display or agent mode```<br>```display_mode = True```<br>```agent_mode = True``` |

Source: compiled by the author.

**Conclusion**

The results of the dissertation research are as follows:

1) Patterns in the development of autonomous AI systems for corporate management have been identified.

2) The taxonomy of autonomous AI systems for corporate management is proposed and substantiated. The emergence of hybrid systems resulting from the unification of multifunctional digital factories and personalised virtual systems or humanoid robots is substantiated.

3) The reference model for the development and implementation of autonomous AI systems is proposed and substantiated based on the synthesis of computational law, a dedicated operational context, controlled generation of synthetic data, game theory (used to calculate the strategy for achieving goals by an AI system), explainable AI technologies and machine learning algorithms.

4) The necessity is substantiated, a methodology for developing algorithmic law for autonomous AI systems for managing corporations is proposed and tested.

5) The necessity is substantiated, a methodology for creating a dedicated operational context for autonomous AI systems for corporate management is proposed and tested.

6) The methodology for training autonomous AI systems for corporate management based on synthetic data has been proposed and tested.

7) The methodology for calculating the strategy of autonomous AI systems for managing corporations based on game theory is proposed and tested.

8) The methodology for developing an interface for autonomous AI systems for corporate management has been proposed and tested.

9) The continuous process of making legitimate and ethical management decisions by autonomous AI systems that combines computational law, dedicated operational context, controlled generation of synthetic data, game theory, explainable AI technologies, and machine learning algorithms is presented.

With the rapid development of artificial intelligence technologies, various aspects of the interaction of technical (AI systems) and social (groups of people) systems must receive a completely new understanding. Autonomous AI systems can mean both a great blessing and a great evil for modern human civilization. As David Chalmers rightly notes, singularity can have unprecedented consequences [11]. However, very few researchers are studying the singularity of even machine artificial intelligence [11]. It seems appropriate for companies producing autonomous AI systems, in parallel with scientific research and their implementation, to simultaneously conduct research in the field of a future social contract that will allow technical and social systems work effectively for the benefit of human civilization.

The path that regulators in many countries have chosen, namely, imposing restrictions on the development and deployment of autonomous AI systems, is not a sustainable strategy in the long term. As Nick Bostrom notes, considering the creation of machine superintelligence, restrictions would be rather meaningless: development may well proceed anyway, either because people do not regard the gradual displacement of biological humans by machines as necessarily a bad outcome, or because such strong forces (driven by short-term profit, curiosity, ideology, or the desire for the possibilities that superintelligence might bring to its creators ) are so active that a collective decision to prohibit new research in this area cannot be reached and successfully implemented [161].

The option of doing nothing and waiting, hoping that the superintelligence will decide everything itself, is also an option. But since the transition to superintelligence will not be instantaneous, this approach will not soften the difficulties of the transition period. One could consider the experience of establishing control over atomic research, when at the international level, under the auspices of the UN, it was possible to establish effective and reasonable use of scientific developments in this area. We have very recent experience of control over global mass phenomena - control over the spread of COVID-19. In the worst case, humanity will have to apply measures developed to control epidemics and pandemics - strict restrictions, lockdowns, special protective equipment. It cannot be said that this method is the best option, which is also possible only until the mass phenomenon has spread very widely.

The transition of modern society to the era of autonomous agents is already underway, and will accelerate with each passing year. The task of not only researchers, but also of every person who looks to the future with hope and sees prospects for effective interaction between biological and artificial intelligence, is to prepare as well as possible for the approaching singularity. It is important to develop approaches at the global level, but the basis of the global approach should be personal understanding. At present, we can say that the personal expectations of users from autonomous AI systems are based more on the ability to make decisions based on big data than on the paradigm of algorithmic leadership (figure 13).

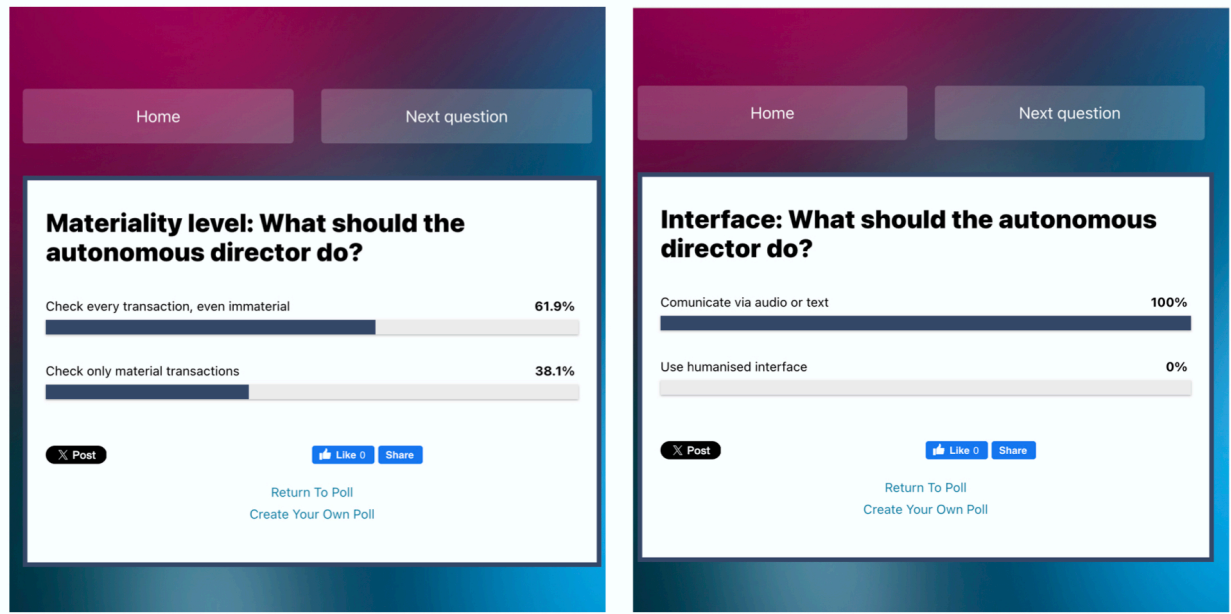

Figure 13 — User preferences for the functionality of the digital director. Source: compiled by the author based on materials from [162].

Numerous aspects of the interaction of technical and social systems in the field of corporate management are the subject of further research.

**Glossary of Terms**

| | |
|---|---|
| **Balance for negative class** | — subjects constituting the negative class from both the protected and unprotected groups must also have the same average predicted probability score S. For example: $E(S \mid Y = 0, Gender = m) = E(S \mid Y = 0, Gender = f)$ |
| **Balance for positive class** | — subjects constituting the positive class from both the protected and unprotected groups must also have the same average predicted probability score S. For example: $E(S \mid Y = 1, Gender = m) = E(S \mid Y = 1, Gender = f)$ |
| **False negative error rate balance / or equal opportunity** | — a classifier satisfies this definition if both the protected and unprotected groups have equal FNR (false negative rate) – the probability that a subject in the positive class will have a negative predictive value. For example: $P(d = 0 \mid Y = 1, Gender = m) = P(d = 0 \mid Y = 1, Gender = f)$ |
| **Positive error rate balance / or predictive equality** | — a classifier satisfies this definition if both the protected and unprotected groups have equal FPR (false positive rate) – the probability that a subject in the negative class will have a positive predictive value. For example: $P(d = 1 \mid Y = 0, Gender = m) = P(d = 1 \mid Y = 0, Gender = f)$ |
| **Non-cooperative games** | - games in which players do not enter into any agreements with each other. |
| **Benchmarking** | — a statistical test of discrimination that compares the degree to which privileged and underprivileged groups are treated favourably. For example: $P(d = 1 \mid Gender = m) = P(d = 1 \mid Gender = f)$ |
| **Top-k selection** | — is a tuning parameter of a large language model that controls the diversity of text by limiting the number of words to the top-k most probable words, regardless of their percentage probabilities. |
| **Top-p selection** | — is a tuning parameter of a large language model that controls the richness of text by limiting the number of words the model can select based on their probabilities. |
| **True negative rate (TNR)** | — the proportion of negative cases correctly predicted to fall into the negative class, out of all actual negative cases $$TNR = \frac{TN}{FP + TN}$$ |
| **True positive rate (TPR)** | — the proportion of positive cases, correct predicted to be in the positive class from all actual positive cases $$TPR = \frac{TP}{TP + FN}$$ |
| **False negative rate (FNR)** | — the proportion of positive cases incorrectly predicted to be in the negative class, out of all actual positive cases $$FNR = \frac{FN}{TP + FN}$$ |
| **False positive rate (FPR)** | — the proportion of negative cases that were incorrectly predicted to belong to the positive class, out of all actual negative cases $$FPR = \frac{FP}{FP + TN}$$ |
| **Law of large numbers** | — in probability theory, the law of large numbers is a mathematical law that states that the mean of results obtained from a large number of independent random samples converges to the true value if one exists (or, given a sample of independent and identically distributed values, the sample mean converges to the true mean). |
| **Protected attribute** | — an attribute in relation to which non-discrimination must be established. |
| **Non-zero sum games** | — the win of one player does not mean the loss of another, that is, the winnings of each player are calculated independently, or interests may coincide, or one of the players may simply not have any goals. |
| **Zero-sum games** | — a situation when the gain of one player is equal to the loss of the other. |
| **True negative (TN)** | — a case where the predicted and actual results belong to the class of negative ones. |
| **True positive (TP)** | — the case where the predicted and actual results are both in the positive class. |
| **Well-calibration** | — for any predictable probability estimate S, subjects in both the protected and unprotected groups must not only have equal probability of actually belonging to the positive class, but this probability must be equal to S. For example: $P(Y = 1 \mid S = s, Gender = m) = P(Y = 1 \mid S = s, Gender = f) = s$ |
| **Cooperative games** | - games in which players can enter into agreements in order to increase their winnings. |
| **Finite games** | — in finite games, players have a finite number of possible strategies. |
| **Counterfactual fairness** | — a causal graph is counterfactual if the predicted outcome d in the graph does not depend on the descendant of the protected attribute G. |
| **False discovery rate (FDR)** | — the proportion of negative cases incorrectly predicted to be in the positive class, out of all predicted positive cases $$FDR = \frac{FP}{TP + FP}$$ |
| **False omission rate (FOR)** | — the proportion of positive cases incorrectly predicted to belong to the negative class out of all predicted negative cases $$FOR = \frac{FN}{TN + FN}$$ |
| **Wald criterion** | — allows you to select the largest element of the payoff matrix from its minimum possible elements ($W = max(min[h_{ji}])$). |
| **Hurwitz criterion** | — is intended to select some average element of the profitability matrix, which differs from the extreme states – from the minimum and maximum elements ($H = max(\lambda*max[a_{ij}] + (1-\lambda)*min[a_{ij}])$). |
| **Savage criterion** | — is designed to select the maximum element of the risk matrix from its minimum possible elements ($S = min(max[r_{ij}])$). |
| **False negative (FN)** | — a case that is predicted to belong to the negative class, whereas the actual outcome belongs to the positive class. |
| **False positive (FP)** | — a case that is predicted to belong to the positive class when the actual outcome belongs to the negative class. |
| **Interquartile range** | — a measure of statistical dispersion, which represents the spread of data (defined as the difference between the 75th and 25th percentiles of the data) - $IQR = Q_3 - Q_1$. |
| **Unprivileged group** | — subsets of individuals in a dataset who have historically been disadvantaged or faced discrimination in social contexts based on certain protected attributes. |
| **Accuracy equality** | — a classifier satisfies this definition if both the protected and unprotected groups have the same prediction accuracy – the probability that an individual from a positive or negative class will be assigned to the corresponding class. For example:<br><br>$$P(d = Y, Gender = m) = P(d = Y, Gender = f)$$ |
| **Negative predictive value (NPV)** | — the proportion of negative cases correctly predicted as belonging to the negative class out of all predicted negative cases<br><br>$$NPV = \frac{TN}{TN + FN}$$ |
| **No unresolved discrimination** | — a causal graph has no unresolved discrimination if there is no path from a protected attribute G to a predicted outcome d except through the resolver variable. |
| **No proxy discrimination** | — a causal graph is free from proxy discrimination if there is no path from a protected attribute G to a predicted outcome d that is blocked by a proxy variable. |
| **Positive predictive value (PPV)** | — the proportion of positive cases correctly predicted to fall into the positive class out of all predicted positive cases<br><br>$$PPV = \frac{TP}{TP + FP}$$ |
| **Predictive parity / or outcome test** | — a classifier satisfies this definition if both the protected and unprotected groups have the same positive predictive value (PPV) — the probability that a subject with a positive predictive value actually belongs to the positive class. For example:<br><br>$P(Y = 1 \mid d = 1, Gender = m) = P(Y = 1 \mid d = 1, Gender = f)$ |
| **Percentile** | — a measure in which the percentage of total values is equal to or less than that measure. |
| **Utility** | — records the preferences of players regarding various game outcomes in terms of real numbers. |
| **Privileged group** | — subsets of individuals in a dataset who have historically received more favourable treatment or had advantages in social contexts based on certain protected attributes. |
| **Causal discrimination** | — a classifier satisfies this definition if it produces the same classification for any two subjects with exactly the same attributes X. For example: $(X_{female} = X_{male} \land G_{female}\,! = G_{male}) \rightarrow d_{female} = d_{male}$ |
| **Treatment equality** | — a classifier satisfies this definition if both protected and unprotected groups have the same ratio of false negatives to false positives. For example: $$\frac{FN}{FP} male = \frac{FN}{FP} female$$ |
| **Equal acceptance rate** | — equal acceptance rates (proportion of positive decisions) for privileged and protected groups of people in a binary classification. For example: $P(d = 1 \,|\, Gender = m) = P(d = 1 \,|\, Gender = f)$ |
| **Signaling** | - providing reliable signals to transmit your private information to other players. |
| **Fair inference** | — this definition classifies paths in the causal graph as legitimate or illegitimate. The causal graph satisfies the notion of fair inference if there are no illegitimate paths from G to d. |
| **Test fairness / or calibration / or matching conditional frequencies** | — a classifier satisfies this definition if , for any predicted probability value S, subjects in both the protected and unprotected groups have the same probability of actually belonging to the positive class . This definition is similar to predictive parity, except that it takes into account the proportion of correct positive predictions for any value of S. For example: $P(Y = 1 \,|\, S = s, Gender = m) = P(Y = 1 \,|\, S = s, Gender = f)$ |

| | |
|---|---|
| **Fairness through unawareness** | — a classifier satisfies this definition if no explicitly sensitive attributes (X) are used in the decision-making process. For example: $$X_i = X_j \rightarrow d_i = d_j$$ |
| **Fairness through awareness** | — fairness is determined by the principle that similar individuals should have similar classifications. The similarity of individuals is determined using a distance metric; to ensure fairness, the distance between the output distributions of individuals should be no greater than the distance between individuals. |
| **Statistical parity** | — a property in which the demographic data of individuals who have received a positive (or negative) classification are identical to the demographic data of the population as a whole. For example: $$P(d = 1 \mid Gender = m) = P(d = 1 \mid Gender = f)$$ |
| **Statistical parity difference** | - compares the percentage of favourable outcomes for control groups with the reference groups. |
| **Temperature** | — a parameter of a large language model that controls the randomness or creativity of the model's output (higher temperature makes the output more varied and unpredictable, while lower temperature makes it more focused and predictable). |
| **Mandelbrot's Fractal Theory** | — fractals are a form of geometric repetition in which successively smaller copies of a pattern are nested within one another, so that the same complex shapes appear repeatedly. Mandelbrot believed that such irregular shapes were in many ways more natural than the artificially smooth objects of traditional geometry, and that they could have wide applications. |
| **Equalized odds / or conditional procedure accuracy equality / or disparate mistreatment** | — a classifier satisfies the definition if the protected and unprotected groups have equal TPR and equal FPR (this is equivalent to combining the conditions for determining the balance of false positive errors and the balance of false negative errors). For example : <br><br> $P(d = 1 \mid Y = i, Gender = m) = P(d = 1 \mid Y = i, Gender = f), i \in 0,1$ |
| **Conditional use accuracy equality** | — combines two conditions : equal PPV and NPV — the probability of subjects with positive predictive value to actually belong to the positive class and the probability of subjects with negative predictive value to actually belong to the negative class. For example: <br><br> $(P(Y = 1 \mid d = 1, Gender = m) = P(Y = 1 \mid d = 1, Gender = f)) \wedge$ <br> $(P(Y = 0 \mid d = 0, Gender = m) = P(Y = 0 \mid d = 0, Gender = f))$ |
| **Conditional statistical parity** | — subjects in both the protected and unprotected groups have the same probability of being assigned to the positive predicted class , controlling for a limited set of legal factors. For example: <br><br> $P(d = 1 \mid L = l, Gender = m) = P(d = 1 \mid L = l, Gender = f)$ |
| **Central Limit Theorem** | — in probability theory, the central limit theorem states that, under appropriate conditions, the distribution of a normalized version of the sample mean converges to the standard normal distribution. |
| **LIME (Local Interpretable Model-agnostic Explanations)** | — is a method that approximates any black-box machine learning model with a local interpretable model to explain each individual prediction. |

# List of figures

# List of tables

**Appendix A**

The synthetic dataset for detecting subsidiary value manipulation using Enron as a case study

| Dataset | URL |
| --- | --- |
| Enron emails synthetic dataset | https://github.com/iboard-project/synthetic-data/tree/main/dataset |

**Appendix B**

The prototype model for making legitimate and ethical decisions by autonomous artificial intelligence systems in corporate management

| Description | URL |
| --- | --- |
| Prototype model | https://github.com/iboard-project/prototype |